\documentclass[sigconf]{acmart}
\usepackage[explicit]{titlesec}
\usepackage{amsmath}
\usepackage{amsfonts}
\usepackage{algorithmic}
\usepackage{graphicx}
\usepackage{textcomp}
\usepackage{xcolor}
\usepackage{amsthm}
\usepackage{subfigure}
\usepackage{colortbl}
\usepackage{soul}  
\usepackage{enumitem}
\usepackage{xcolor}
\usepackage{fancyhdr}
\usepackage{pifont}
\definecolor{skyblue}{rgb}{0., 0.72, 0.92}
\usepackage[outline]{contour}
\usepackage[linesnumbered,ruled,vlined]{algorithm2e}
\SetKwInOut{Input}{Input}
\SetKwInOut{Output}{Output}
\SetCommentSty{mycommfont}

\usepackage{multirow}
\usepackage{booktabs}
\usepackage{tabularx, booktabs}

\usepackage{balance}
\usepackage[utf8]{inputenc} 
\usepackage[T1]{fontenc}    
\usepackage{hyperref}       
\usepackage{url}            
\usepackage{booktabs}       
\usepackage{amsfonts}       
\usepackage{nicefrac}       
\usepackage{microtype}      
\usepackage{multirow}
\usepackage{rotating}
\usepackage{bbding}
\usepackage{subfigure}
\usepackage{listings}

\usepackage{xcolor}
\usepackage[linesnumbered,ruled,vlined]{algorithm2e}
\SetKwInOut{Input}{Input}
\SetKwInOut{Output}{Output}

\SetCommentSty{mycommfont}
\SetKwInput{KwInput}{Input}            
\SetKwInput{KwOutput}{Output}
\definecolor{answercolor}{RGB}{240, 240, 240}

\theoremstyle{definition}

\def\BibTeX{{\rm B\kern-.05em{\sc i\kern-.025em b}\kern-.08em
    T\kern-.1667em\lower.7ex\hbox{E}\kern-.125emX}}
    
\AtBeginDocument{%
  \providecommand\BibTeX{{%
    \normalfont B\kern-0.5em{\scshape i\kern-0.25em b}\kern-0.8em\TeX}}}

\begin{document}

\title{Active Code Learning: Benchmarking Sample-Efficient Training of Code Models}

\author{
Qiang Hu$^{1}$,
Yuejun Guo$^{2}$,
Xiaofei Xie$^{3}$
Maxime Cordy$^{1}$,
Lei Ma$^{4,5}$,
Mike Papadakis$^{1}$,
and Yves Le Traon$^{1}$
\\ 
\normalsize{
$^1$University of Luxembourg, Luxembourg \quad \\
$^2$Luxembourg Institute of Science and Technology, Luxembourg \quad \\
$^3$Singapore Management University, Singapore \quad \\
$^4$The University of Tokyo, Japan \quad \\
$^4$ University of Alberta, Canada \quad \\
}
}
\renewcommand{\shortauthors}{Hu et al.}
\begin{abstract}

The costly human effort required to prepare the training data of machine learning (ML) models hinders their practical development and usage in software engineering (ML4Code), especially for those with limited budgets. Therefore, efficiently training models of code with less human effort has become an emergent problem. Active learning is such a technique to address this issue that allows developers to train a model with reduced data while producing models with desired performance, which has been well studied in computer vision and natural language processing domains. Unfortunately, there is no such work that explores the effectiveness of active learning for code models. In this paper, we bridge this gap by building the first benchmark to study this critical problem - active code learning. Specifically, we collect 11 acquisition functions~(which are used for data selection in active learning) from existing works and adapt them for code-related tasks. Then, we conduct an empirical study to check whether these acquisition functions maintain performance for code data. The results demonstrate that feature selection highly affects active learning and using output vectors to select data is the best choice. For the code summarization task, active code learning is ineffective which produces models with over a 29.64\% gap compared to the expected performance. Furthermore, we explore future directions of active code learning with an exploratory study. We propose to replace distance calculation methods with evaluation metrics and find a correlation between these evaluation-based distance methods and the performance of code models.

\end{abstract}

\maketitle

\section{Introduction}
\label{sec:intro}
Using machine learning (ML) to help developers solve software problems~(ML4Code)~\cite{allamanis2018survey} has been a hot direction in both software engineering (SE) and ML communities in recent years. Deep learning (DL), one of the advanced ML techniques, has achieved great success in multiple software tasks, such as code summarization~\cite{ijcai2018p314}, code clone detection~\cite{8094426}, and vulnerability detection~\cite{zhou2019devign}. Typically, preparing a code model involves two main steps: first, building a pre-trained model that learns general code information; second, fine-tuning this model using datasets that target a specific downstream task. Both components contribute to the success of ML4Code and are still under exploration for further improving the performance of code models.

Generally, pre-trained code models can be easily accessed from open resources, e.g., Hugging Face~\cite{wolf-etal-2020-transformers}, or built by using self-supervised learning without data labeling effort. This means that for developers planning to use code models, the first step of pre-trained model preparation is not challenging and can be fully automated. However, collecting datasets to fine-tune pre-trained models is not easy. The main reason is that the general fine-tuning process follows the procedure of fully-supervised learning, which requires carefully labeled training data. Unfortunately, the data labeling process is time-consuming and labor-intensive \cite{zhou2019devign}.

To alleviate the aforementioned heavy effort of training data labeling, active learning~\cite{settles2009active} is used to enable sample-efficient model training in other famous fields, e.g., computer vision~\cite{sener2018active} and natural language processing~\cite{settles2008analysis}. The key idea of active learning is to iteratively select a subset of training data to label and use them to train the model. The existing studies~\cite{hu2021towards} have shown that labeling only a few (less than 10\%) training data can train a model with similar performance as the model trained by using the entire training data. In this way, the labeling effort can be significantly reduced and made flexible to a fixed budget. However, despite being well-studied in many application domains, the usefulness of active learning in ML4Code is still unknown. Researchers mainly focus on designing new model architectures, proposing novel code representation methods, and studying more software tasks. The study of how to lighten the model training cost is missed. There is a need to provide a benchmark to support the exploration of this important problem.

In this paper, we aim to bridge this gap and build a benchmark to study how active learning can help us efficiently build code models -- active code learning.  We collect acquisition functions~(i.e., active learning methods) that can be used for code data from existing studies~\cite{hu2021towards, weiss2022simple}. In total, we implement 11 acquisition functions~(including random selection) that can be divided into two groups, output-uncertainty-based functions~(5 functions) and clustering-based functions~(5 functions) in our benchmark. Then, based on these collected acquisition functions, we conduct experiments on four datasets~(including classification tasks and non-classification tasks) and two famous pre-trained code models to answer the following research questions:  

\textbf{RQ1: What features should be used for clustering-based acquisition functions? } Feature selection~\cite{liu2005toward} is an important problem for clustering methods. However, it is unclear which features should be used for clustering-based methods in active code learning. Our first study is to explore how different features affect the effectiveness of active code learning. Here, we consider three types of features, code tokens, code embedding vectors, and model output vectors. \textit{Findings:} The results show that clustering-based methods are sensitive to the used features. Interestingly, in more than half of cases (62.5\%), output vectors based achieves significantly better results than code token and code embedding-based methods. 

\textbf{RQ2: How do acquisition functions perform on code models?} After determining features to use, we compare all the acquisition functions across different tasks. \textit{Findings:} Firstly, contrary to the previous study~\cite{weiss2022simple}, which suggests that simple techniques perform better for active learning,  we found that clustering-based methods consistently outperform simple uncertainty methods in our considered binary classification code task~(Clone detection). Secondly, unlike prior research~\cite{hu2021towards, weiss2022simple}, which shows that only a small set of training data (less than 10\%) is sufficient to produce good models, our results on code summarization tasks indicate that existing methods are not yet capable of achieving this goal. There is at least a 30\% performance gap between the models trained by active learning (with 10\% data) and the models trained using the entire training data. In total, in the first two studies, we trained 5200 models considering each labeling budget. The training process takes more than 3200 GPU hours. 

\textbf{Exploratory study.} Additionally, using our benchmark, we conduct a study to further explore potential directions for proposing new acquisition functions. We focus on clustering-based methods since existing methods can not perform well on non-classification tasks. Concretely, due to clustering-based methods tending to select diverse data (data have a bigger distance to each other), we first check if there is a correlation between the distance of selected data and the accuracy of the trained model. Then, we propose a novel view to consider the distance of code for active learning -- \textit{using evaluation metrics~(e.g., CodeBERTScore) as distance calculation methods}. Based on the evaluation, we found that, for non-classification code tasks, 1) there is no correlation between the distance~(calculated by Cosine similarity and Euclidean distance) of selected data to each other and the accuracy of models. 2) There is a weak correlation between the evaluation metrics-based distance of selected data to each other and the accuracy of models, indicating that future proposed methods can be based on evaluation metrics-based distance.

To summarize, the main contributions of this paper are:
\begin{itemize}[leftmargin=*]
\item This is the first work that builds a benchmark for sample-efficient training for code models. 

\item Based on our benchmark and empirical study, we found that multiple findings from previous works~\cite{hu2021towards, weiss2022simple} on image and text data are inapplicable to code data.

\item We design a new strategy that uses evaluation metrics to compute the distance between code pairs to support future research when proposing new acquisition functions.
\end{itemize}


\section{Background}
\label{sec:background}

\subsection{Active Learning}

\begin{figure}[]
	\centering
	\includegraphics[width=0.475\textwidth]{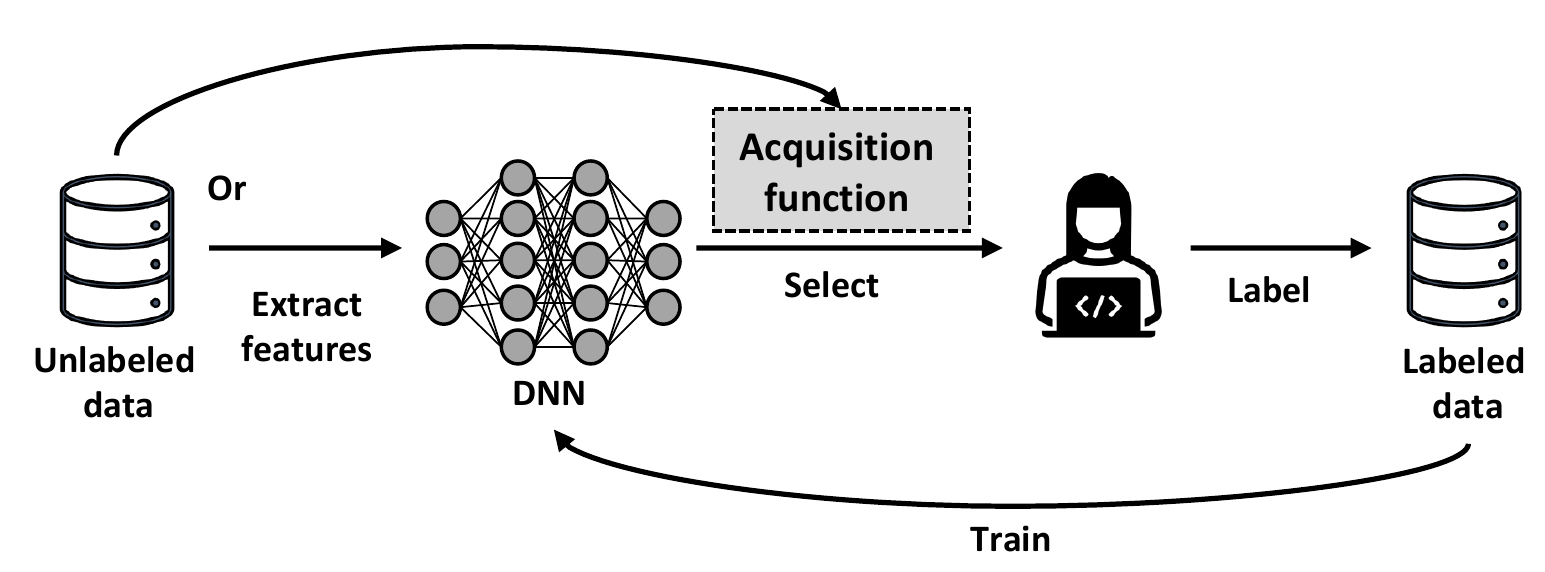}
	\caption{Overview of active learning.}
	\label{fig:al}
\end{figure}

Active learning is a well-known technique to enable sample-efficient model training. Figure~\ref{fig:al} depicts the overview workflow of active learning. Given an unlabeled dataset and a model under training, the first step of active learning is to choose the features used for conducting data sampling. Generally, two types of features can be used, 1) the data features itself~(e.g., image pixels, and code tokens), and 2) features extracted from the model~(e.g., output probabilities, and code embeddings). After obtaining the features, acquisition functions are used to select the most valuable data for labeling. Here, important means the model can learn more information from this kind of data and achieve better performance on test data. Finally, developers label these selected data and use them to train the model. In this way, developers can train a model under a fixed labeling budget. When a new budget is allowed, developers repeat this process and further enhance the trained model.

The acquisition function is the most important part of active learning which decides the quality of labeled training data. Existing acquisition functions can be roughly divided into two groups, output-uncertainty-based functions, and clustering-based functions. Output-uncertainty-based functions obtain the model predictions of the unlabeled data first and then use some uncertainty metrics~(Entropy of output probabilities) to rank these data and select the most uncertain ones for training. The intuition behind this type of function is that uncertain data are usually close to the decision boundaries of the model, thus, training the model using these data can force the model to build clear boundaries and improve its correctness. Clustering-based functions assume that the models have better performance if they learn diverse more data information. Thus, people use different distance methods to measure the distance of data to each other and select the central data for model training, e.g., using K-means to do data clustering and then select the center data.

\begin{table*}[]
\centering
\caption{Our considered acquisition functions. Both: classification and non-classification tasks.} 
\label{table:methods}
\resizebox{1\textwidth}{!}{
\begin{tabular}{lcl}
\hline
\textbf{Method Name} & \multicolumn{1}{l}{\textbf{Support Tasks}} & \textbf{Description} \\ \hline
Random & Both & Randomly select a budget of data. \\ 
Least Confidence (LC)~\cite{wang2014new} & Classification & Select the data with the least top-1 probability. \\ 
DeepGini~\cite{feng2020deepgini} & Classification & Select data with minimum Gini impurity $\sum\limits_{i=1}^{N}\left(p_i\left(x\right)\right)^2$. \\
Bayesian Active Learning by Disagreement (BLAD)~\cite{gal2017deep} & Classification & Select data minimum disagreement $\frac{count\left(mode\left(y_x^1,...,y_x^T\right)\right)}{T}$ over $T$ dropouts.\\
Entropy~\cite{wang2014new} & Classification & Select data with the minimum Shannon entropy on output probabilities.\\
Margin~\cite{wang2014new} & Classification &  \begin{tabular}[c]{@{}l@{}}Select data with a minimum difference between the top-1 probability \\ and top-2 probability. \end{tabular} \\
Contrastive Active Learning~(CAL)~\cite{margatina-etal-2021-active} & Classification & \begin{tabular}[c]{@{}l@{}}Select a subset that 1) is far away from the labeled data, \\ 2) included data samples have big output divergence. \end{tabular} \\
K-Means (KM)~\cite{sener2018active} & Both & \begin{tabular}[c]{@{}l@{}}Use the K-means clustering method to divide data into K groups and select \\ the center of each group.\end{tabular} \\
K-Center (KC)~\cite{sener2018active} & Both & Iteratively select data that are far from the labeled data.\\
BADGE~\cite{Ash2020Deep} & Both & Select data by using K-means on the gradient embedding level.\\
Coreset~\cite{sener2018active} & Both & An advanced version of K-Center with a distance upper bound.\\ \hline
\end{tabular}
}
\end{table*}

In this work, we follow the previous works~\cite{hu2021towards, weiss2022simple} and collect and implement 11 acquisition functions for active code learning.  Let $X$ be the training data and $x \in X$ be an input sample. $N$ is the number of classes. $y$ and $y_{x}$ are the ground-true label and predicted labels.  $p_{i}(x)$ represents
the predicted probability of $x$ belonging to the $i$th class. Table~\ref{table:methods} presents these acquisition functions with their brief descriptions. For detailed information on each function, please refer to the original paper.

\subsection{Machine Learning for Code (ML4Code)}

ML4Code~\cite{allamanis2018survey} is a new but hot direction in the current software engineering research field. The basic idea of ML4Code is to train a machine learning model to solve software problems, for example, program repair, vulnerability detection, and code summarization. Machine learning models and training data play key roles in ML4Code. For machine learning models, due to the \textit{naturalness} of software, researchers believe model architectures that are good at handling text data are also promising for solving code data. Thus, the famous code models mainly come from modifying natural language models, e.g., CodeBERT~\cite{feng-etal-2020-codebert}. The training data are usually processed by some code representation techniques to transfer the raw data to a machine-readable format, e.g., transfer the strings of code to a sequence of tokens with integer format. How to design this transformation (i.e., code representation) is another research topic that highly affects the performance of trained models. 

In this work, we focus on the most practical paradigm of ML4Code -- building models of code for specific downstream code tasks by fine-tuning pre-trained code models. Existing studies~\cite{feng-etal-2020-codebert, guo2020graphcodebert} already demonstrated that fine-tuning pre-trained models can achieve better results than training the models from scratch. Here, the pre-trained code model has been trained on multi-language datasets, therefore, they already learned general information of code, e.g., the pre-trained CodeBERT model learns knowledge from six datasets with different programming languages. As a result, developers only need to prepare their dataset for a specific task and use them to fine-tune the pre-trained model.

\section{Benchmark}
\label{sec:bencj}

After collecting massive acquisition functions whose effectiveness on image data and text data has already been demonstrated in previous studies, we plan to study their unknown use case -- if they are still useful for efficient training models of code. As mentioned in Section~\ref{sec:intro}, we have two main questions want to answer, 1) since features used for clustering-based acquisition functions are important and highly related to the final performance of active learning, we explore four each acquisition function, which kind of feature is suitable for conducting code selection. 2) As the previous study~\cite{weiss2022simple} claimed that -- simple methods perform well on active learning,  we want to check if this conclusion still holds on active code learning. Concretely, we compare output-uncertainty-based functions~(i.e., simple methods mentioned in~\cite{weiss2022simple}) with clustering-based functions with carefully chosen features.

\subsection{Study Design}

To address the question of suitable feature selection~(\textbf{RQ1}), we first prepare different versions of acquisition functions based on the features they used. Then, we conduct active learning on different code tasks using these functions and compare the performance of trained models. we can draw conclusions about which features are most suitable for each acquisition function.  Here, we focus on three  types of features from different perspectives, 1) code embeddings, 2) sequence of code tokens, and 3)model outputs vectors. 

\begin{itemize}[leftmargin=*]

\item \textbf{Code Embeddings} It is common to consider code embeddings as the input of clustering methods since code embeddings produced by pre-trained code models can present the general information of code. Code embedding is similar to the image data after image pre-processing. In our study, code embeddings extracted from the fine-tuned code models which can better represent the downstream task are used as the features.

\item \textbf{Sequence of code tokens} Regardless of the code representation techniques, the raw program will be converted into a sequence of integer values. Thus, it is also possible to use these sequences as the inputs of the clustering methods. It is similar to using the pixel numbers of images as inputs. The only difference is that code tokens are often with bigger data space, i.e.,  the image pixels are from 0 to 255, while the range of code tokens depends on the vocabulary size, which is generally much bigger than 255, e.g., the vocabulary size of our used CodeBERT model is 50265.

\item \textbf{Model outputs} Different from the above two features that have been widely explored in other fields and can be easily considered for active code learning, in our study, we propose to use the model outputs as the input features of clustering methods. The intuition behind this idea is that the output can be seen as the \textit{understanding} of the model on this input data which should be useful for data selection. Specifically, for classification tasks, we use the one-hot output probabilities as the features, and for the non-classification tasks, the output vectors produced by decoders are used as the features.

\end{itemize}

For the comparison of each acquisition function ~(\textbf{RQ2}), the main goal is to find the recommended function that has consistently better performance than others in active code learning. At the same time, we also compare the output-uncertainty-based functions and clustering-based functions to determine if the previous findings~\cite{weiss2022simple} hold true in active code learning. Here, based on the first study, we use suitable features as the input for clustering-based acquisition functions and perform code selection.

\subsection{Dataset and Model}

\begin{table*}[]
\centering
\caption{Details of datasets and models. Accuracy (\%) for problem classification, F1-score~(\%) for clone detection, and PPL/BLEU for code summarization.}

\label{table:data}
\begin{tabular}{lccccc}
\hline
\textbf{Task} & \textbf{Dataset} & \textbf{Language} & \textbf{Train/Dev/Test} & \multicolumn{1}{l}{\textbf{CodeBERT}} & \multicolumn{1}{l}{\textbf{GraphCodeBERT}} \\ \hline
Problem classification & Java250 & Java & 62500/-/12500 & 98.10 & 98.49 \\
Clone detection & BigCloneBench & Java & 90102/4000/4000 & 97.15 & 97.05 \\
\multirow{2}{*}{Code Summarization} & CodeSearchNet & JavaScript & 58025/3885/3291 & 3.85/14.34 & 3.79/14.89 \\
 & CodeSearchNet & Ruby & 24927/1400/1261 & 4.04/12.80 & 3.99/13.54 \\ \hline
\end{tabular}
\end{table*}

Table~\ref{table:data} presents the datasets and models we used in the study. We consider three code tasks including a multi-class classification task (problem classification), a binary classification task~(clone detection), and a non-classification task ~(code summarization). 

\begin{itemize}[leftmargin=*]

\item \textbf{Problem classification} is a multi-class classification task. Given a program, the model predicts the target problem that the program solves. We use the dataset JAVA250~\cite{Puri0JZDZD0CDTB21} provided by IBM for this task. JAVA250 contains 250 code problems, e.g., problem: \textit{write a program which prints heights of the top three mountains in descending order.} 

\item \textbf{Clone detection} is a well-studied task in the software engineering field to lighten the effort of software maintenance. The main purpose of this task is to check if two programs are semantically equivalent. Thus, code clone detection is often seen as a binary classification problem. In our study, we use the dataset provided by BigCloneBench~\cite{svajlenko2014towards} which contains a large number of Java code clone pairs. Besides, for the computation friendly, we follow the previous work~\cite{yang2022natural} and only use a subset of data from BigCloneBench. 

\item \textbf{Code summarization} is a common code task for helping developers understand code snippets. Given a program, the model generates natural language comments to describe the functionality of this program. We use two datasets provided by Microsoft~\cite{DBLP:journals/corr/abs-2102-04664} in our study. 

\end{itemize}

For the code models, we focus on pre-trained code models since they achieved much better results than models trained from scratch~\cite{feng-etal-2020-codebert, guo2020graphcodebert}.  We follow the previous work~\cite{yang2022natural} and use two well-known pre-trained code models in our study, CodeBERT~\cite{feng-etal-2020-codebert} and GraphCodeBERT~\cite{guo2020graphcodebert}. Other models can be easily added and evaluated to our provided project. Considering the downstream tasks, the pre-trained model is used as an encoder to generate code embeddings, and a decoder is followed to produce the final results for a specific code problem. For example, a dense layer is used as the decoder to produce the output probabilities of classification tasks. The same as the original works or CodeBERT and GraphCodeBERT, during the fine-tuning, we also fine-tune the parameters of pre-trained encoders to ensure a better performance on downstream tasks. 

\begin{itemize}[leftmargin=*]

\item \textbf{CodeBERT} is a bimodal model that shares a similar architecture with the well-known BERT model in the field of natural language processing. The CodeBERT model is initialized through pre-training with six code datasets featuring various programming languages such as Java and Python. During the training process, programs are transformed into sequences of tokens, which is similar to text data. As a result, CodeBERT is able to learn the semantic information of code data.

\item \textbf{GraphCodeBERT} is a newer pre-trained code model that incorporates both sequences of tokens and data-flow information to learn code knowledge. This allows pre-trained code models to understand the structural information of programs and generate more precise code representations.  In our study, the base model is learned by this combination of code information and fine-tuned for our considered downstream tasks only using the code token information. The reason is that we found in some downstream tasks, adding the data-flow information to fine-tune the model can bring a negative influence on the performance of models, e.g., for the problem classification task, adding data-flow information and without data-flow information, the accuracy of fine-tuned models are  82.30\% and 98.39\%, respectively.  

\end{itemize}

\subsection{Evaluation Metrics}

Our study contains three types of code tasks. For each task, we use the most practical metrics to evaluate the trained models. 

\textbf{Accuracy} on the test data is the basic way to evaluate the performance of multi-class classification models. It calculates the percentage~(\%) of correctly classified data over the entire
input data.

\textbf{F1-score} is a commonly used metric for binary classification problems. It calculates the harmonic mean of the precision and recall scores. Given that, true positive~(TP) represents the number of samples correctly predicted as positive, false positive~(FP) represents the number of samples wrongly predicted as positive, and false negative~(FN) represents the number of samples wrongly predicted as negative, F1-score is calculated as:

$$F1-score=\frac{TP}{TP + \frac{1}{2}\left(FP + FN\right)}$$

\textbf{Perplexity~(PPL)} is a widely used metric to evaluate language models. PPL can be seen as the loss of language models that can record the logs of the training process. A lower PPL score means a better performance of the model. Recently, researchers applied PPL to record to evaluate the code summarization models~\cite{wan2018improving}. Specifically, PPL is calculated by:

$$PPL\left(X\right)=\exp\{-\frac{1}{t}\sum\limits_{i=0}^{t}\log{p_{\theta}}(x_i|x_{<i})\}$$

\noindent where $X$ = ($x_{0}$, $x_{1}$,...,$x_{t}$) is the set of code tokens, and $logp_{\theta}(x_i|x_{<i})$ is the log-likelihood of the $i$th token conditioned on the preceding tokens $x_{<i}$ by the model.

\textbf{BLEU~(Bilingual Evaluation Understudy)} is a metric to evaluate the quality of the generated text to another~(the reference). Simply, BLEU score is calculated by:

$$BLEU=BP\times \exp{\sum\limits_{n=1}^{N}w_{n}\log{p_{n}}}$$
The value of N depends on the used N-gram precision, and the weights $w_{n}$ = N / 4. in our study, 4 is used. $p_{n}$ is the ratio of length n sub-sequences in both the candidate sequence and the reference. BP is the brevity penalty calculated by:

$$BP = \left\{ \begin{array}{rcl}1, \qquad if c > r \\
        e^{1 - r/c}, \qquad if c \leq r \end{array}\right. $$

\noindent where $c$ is the length of the generated sequence and $r$ is the length of the reference sequence.

\subsection{Configurations}

\textbf{Training configuration.} Considering the hyperparameters used for model training, we follow the previous work~\cite{yang2022natural} and use the same batch size and learning rate for each model. All the detailed settings can be found on our project site. For model fine-tuning, we set the training epoch as 10 which is enough to ensure the convergence of models. 

\textbf{Active learning configuration.}We initialize the models~(which is a common setting of active learning that we start from a model with a little knowledge of the datasets) by training them on randomly selected 500 samples. We set the labeling budget as 1\% of the entire training set and do 10 times iterations of active learning. 

\textbf{Acquisition function configuration.} For all clustering-based methods, we follow the previous work~\cite{hu2021towards} and set the number of centers as the labeling budget. For BALD, we set the times of dropout prediction as 20.

\subsection{Implementation and Environment}

The project is based on Python-3.6 and PyTorch-1.10.2 framework. The key implementation of code models is modified from the open source project ~\cite{DBLP:journals/corr/abs-2102-04664}. We adopt acquisition functions  implemented by~\cite{hu2021towards, weiss2022simple} that are used for image data and text data to code data. All the source code can be found on our project site. We conduct all the experiments on a
2.6 GHz Intel Xeon Gold 6132 CPU with an NVIDIA Tesla
V100 16G SXM2 GPU. We repeat all the experiments five times to reduce the influence of randomness and report the average results in the following sections.

\subsection{RQ1: Feature Selection}

\begin{figure*}[h]
    \centering
    \subfigure[Problem Classification, KM]{
    \includegraphics[scale=0.27]{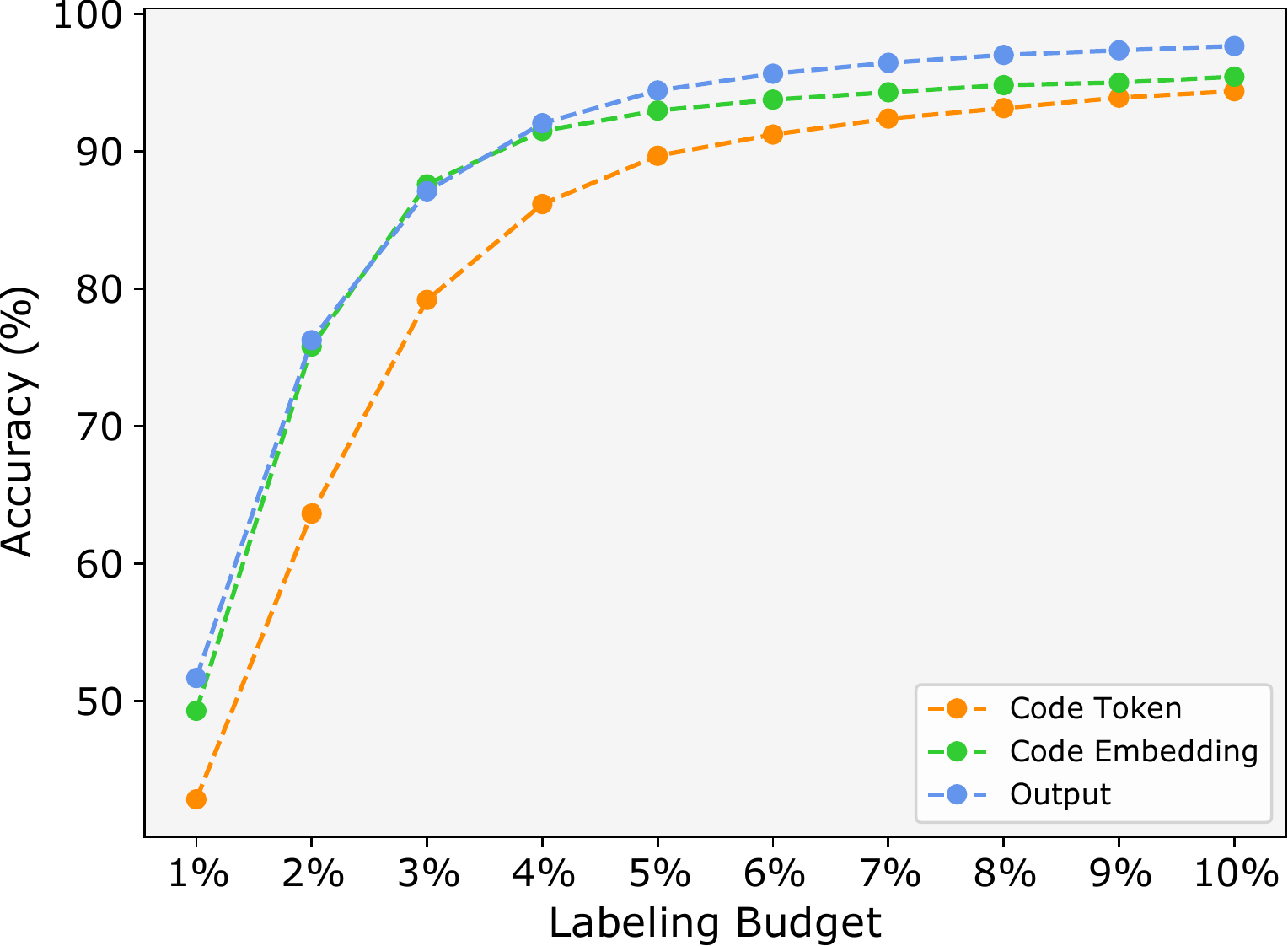}%
    }
    \subfigure[Problem Classification, KC]{
    \includegraphics[scale=0.27]{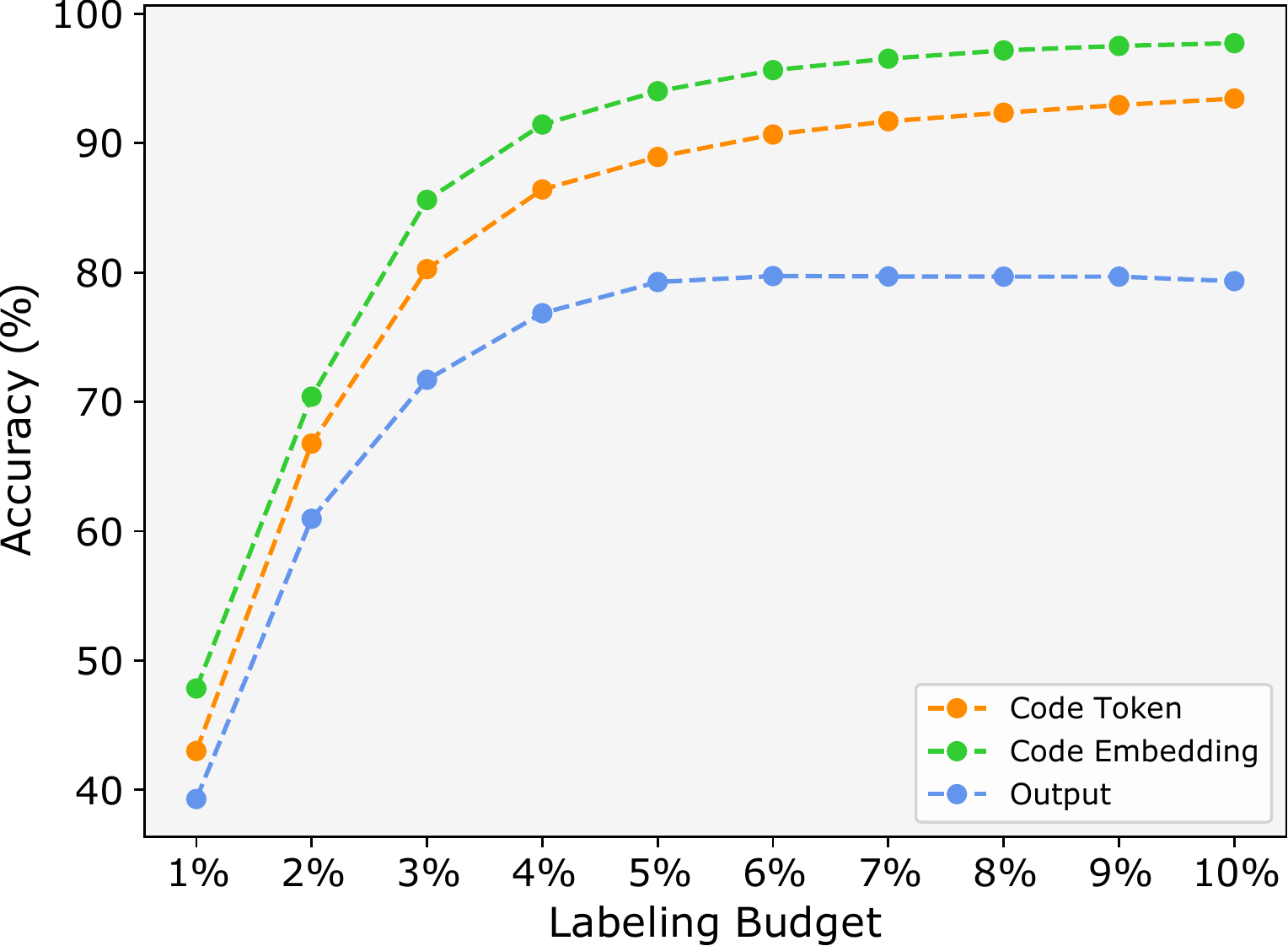}%
    }
    \subfigure[Problem Classification, BADGE]{
    \includegraphics[scale=0.27]{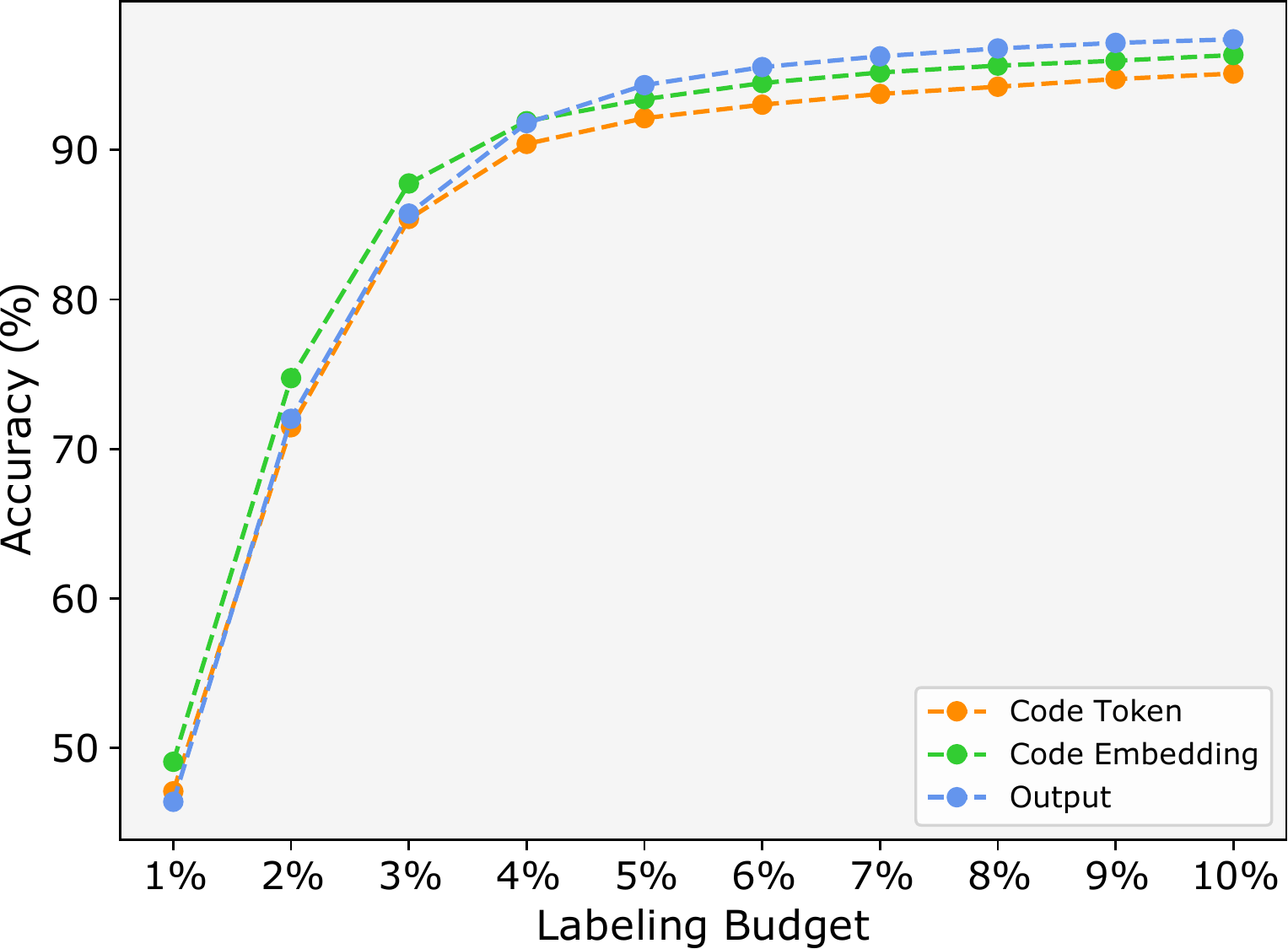}%
    }
    \subfigure[Problem Classification, Coreset]{
    \includegraphics[scale=0.27]{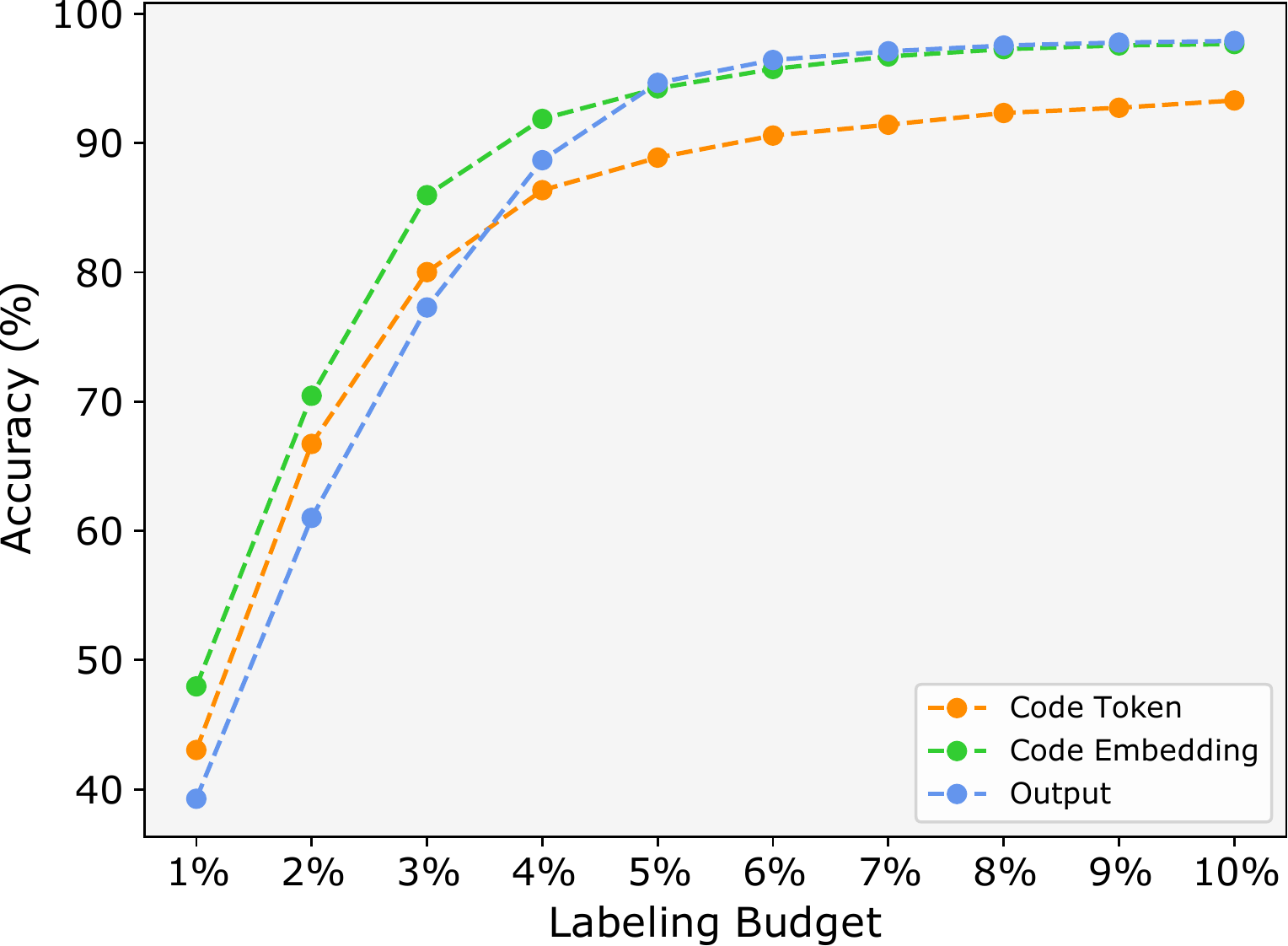}%
    }
    \subfigure[Clone Detection, KM]{
    \includegraphics[scale=0.27]{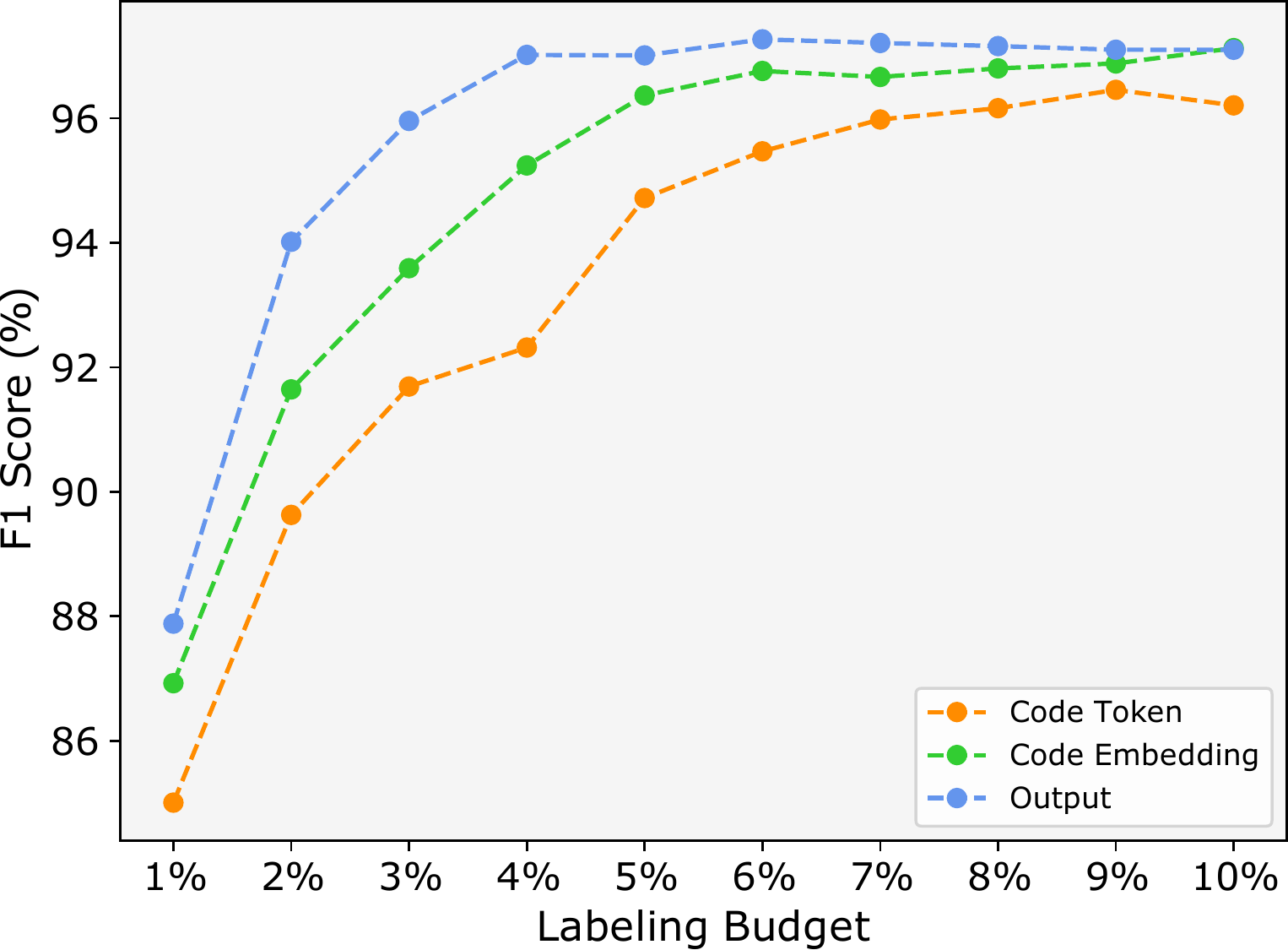}
    }
    \subfigure[Clone Detection, KC]{
    \includegraphics[scale=0.27]{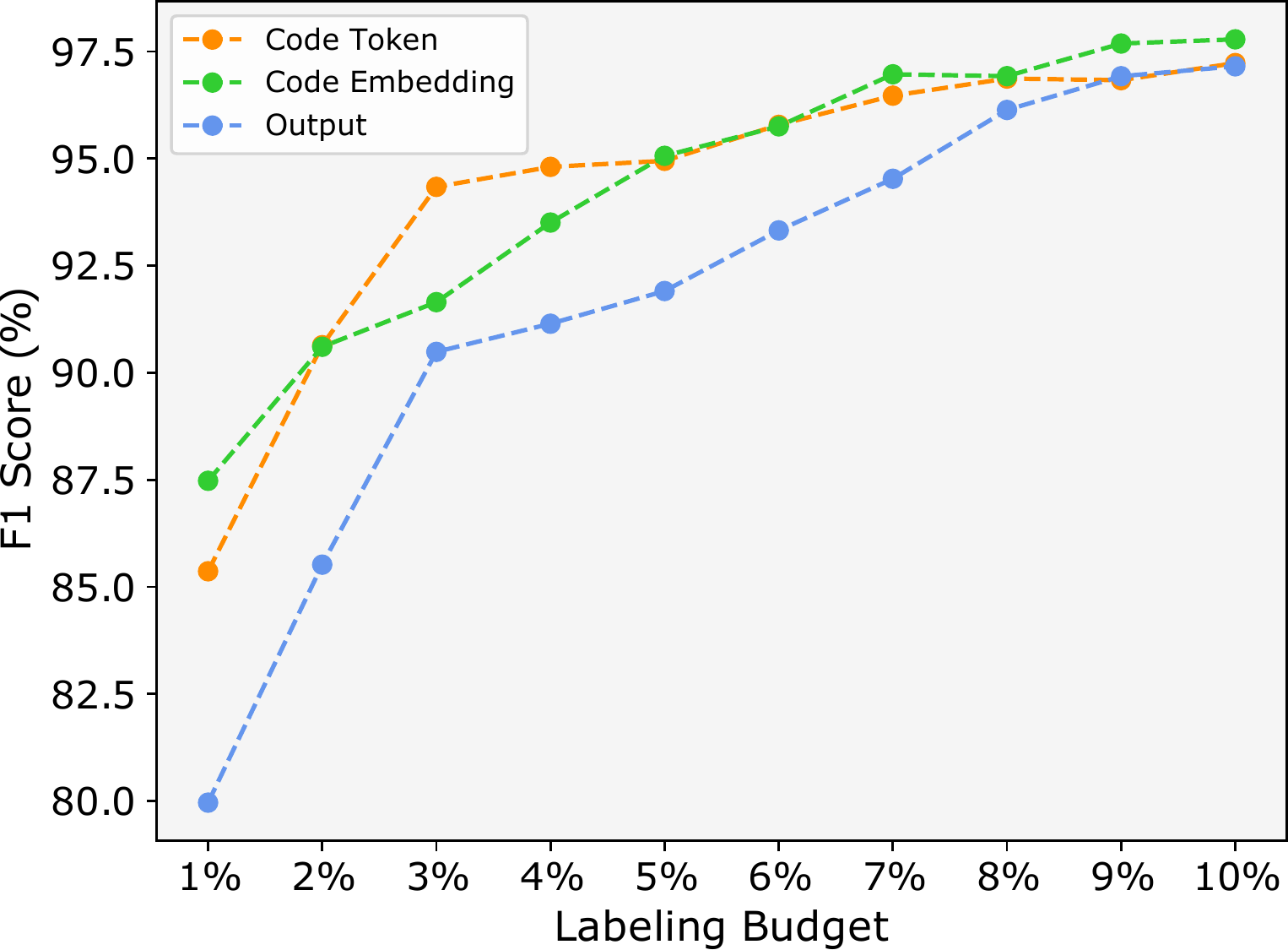}%
    }
    \subfigure[Clone Detection, BADGE]{
    \includegraphics[scale=0.27]{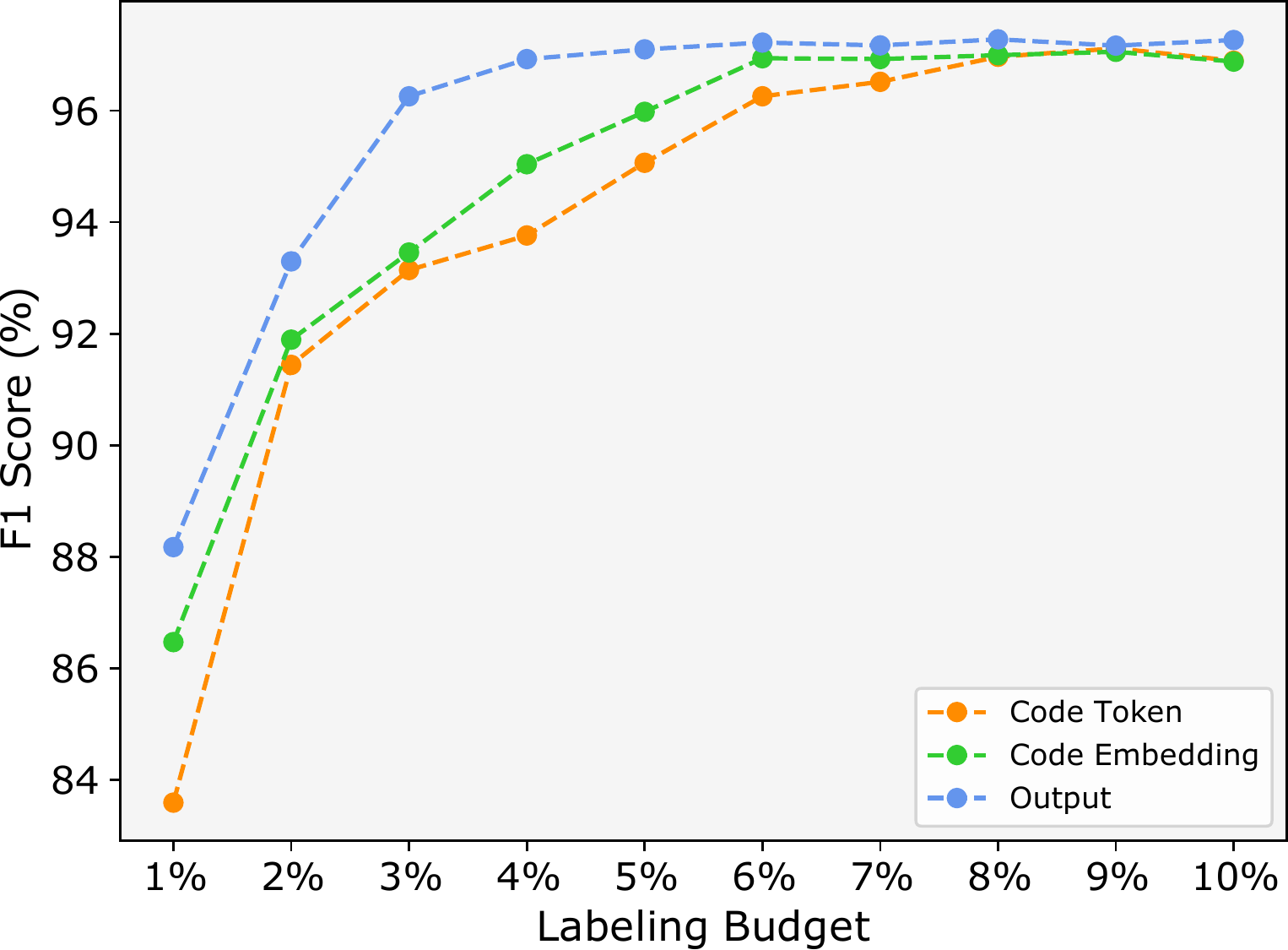}%
    }
    \subfigure[Clone Detection, Coreset]{
    \includegraphics[scale=0.27]{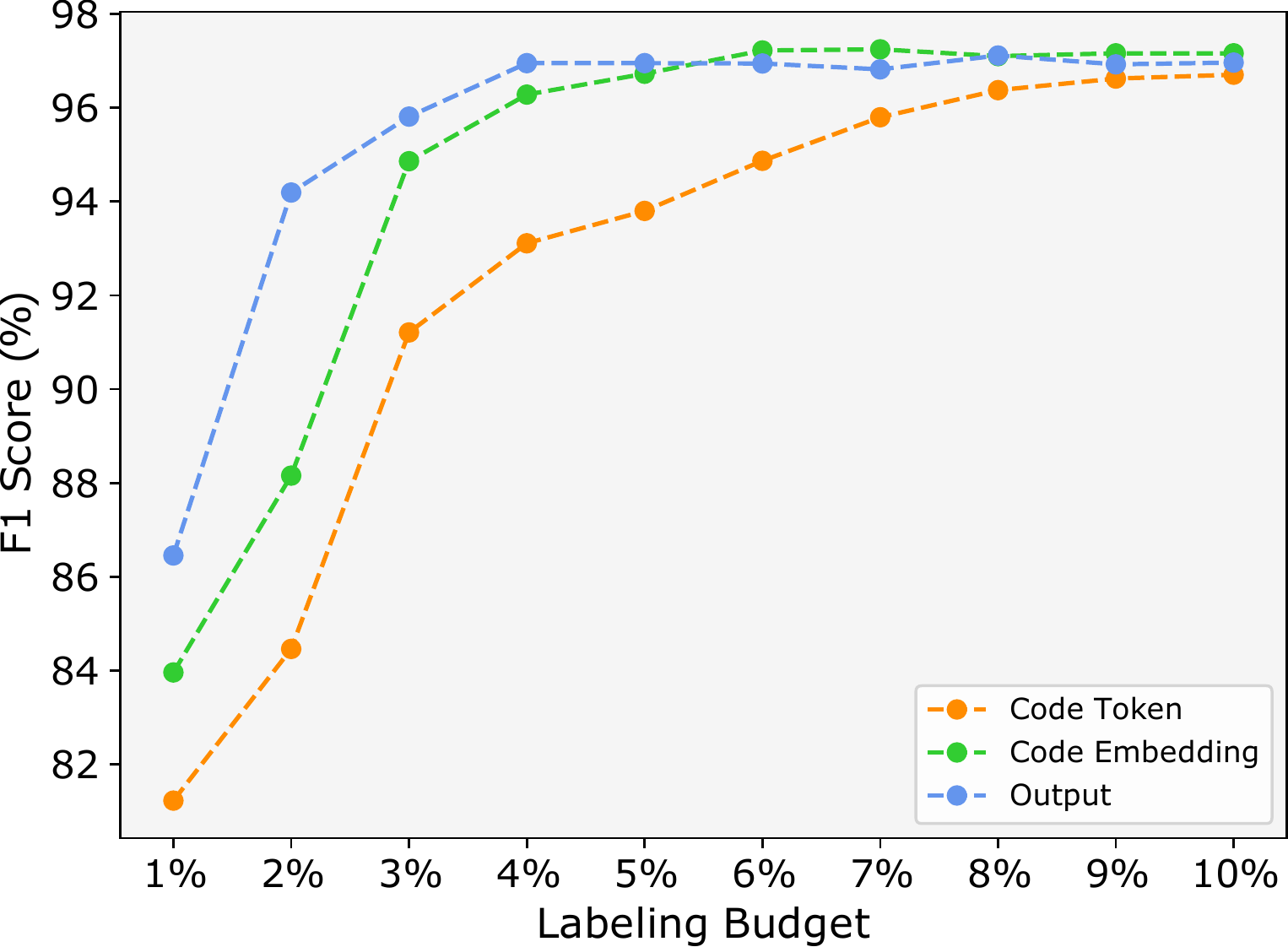}%
    }
    \subfigure[Code Summarization-JS, KM]{
    \includegraphics[scale=0.27]{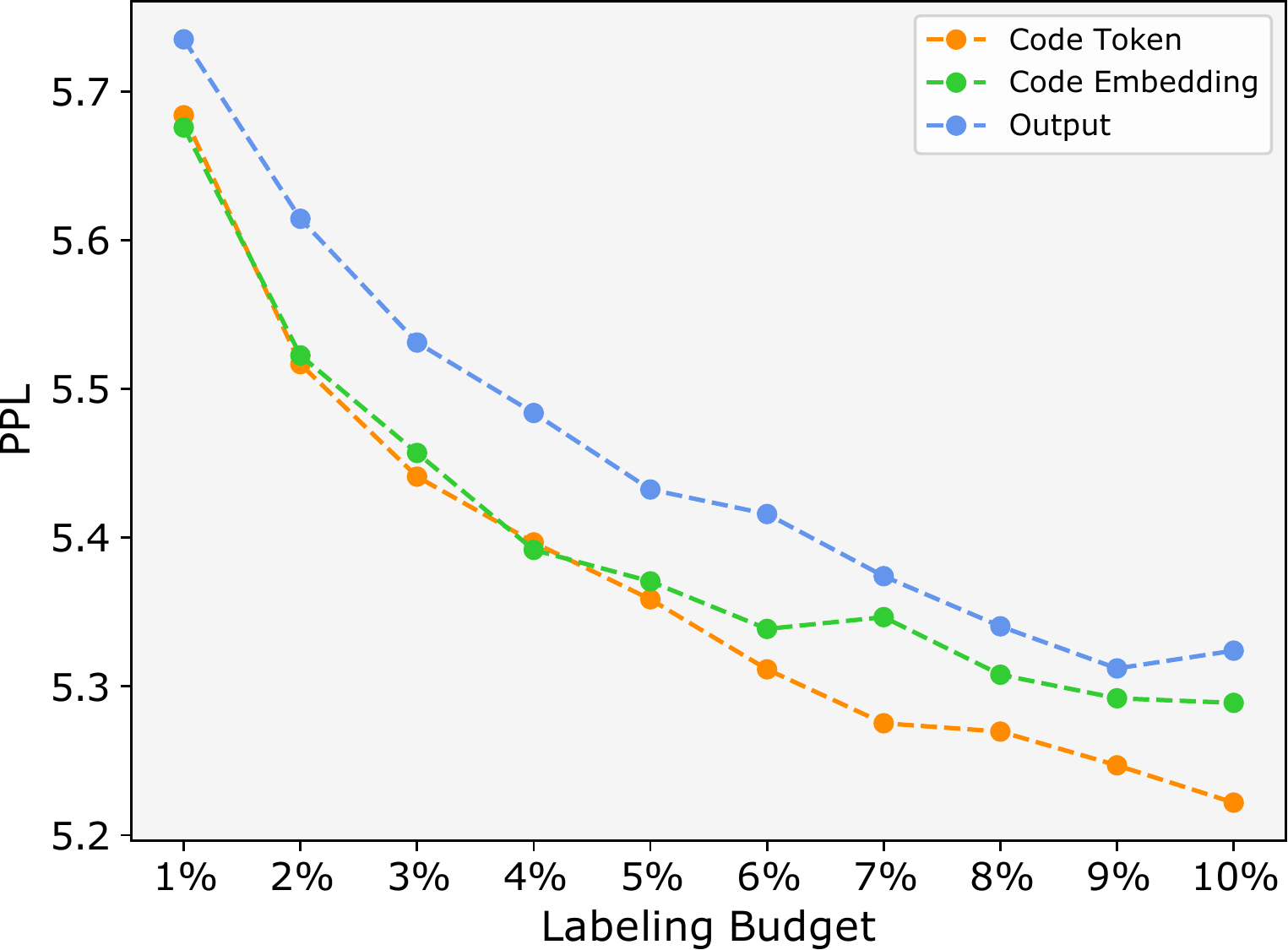}%
    }
    \subfigure[Code Summarization-JS, KC]{
    \includegraphics[scale=0.27]{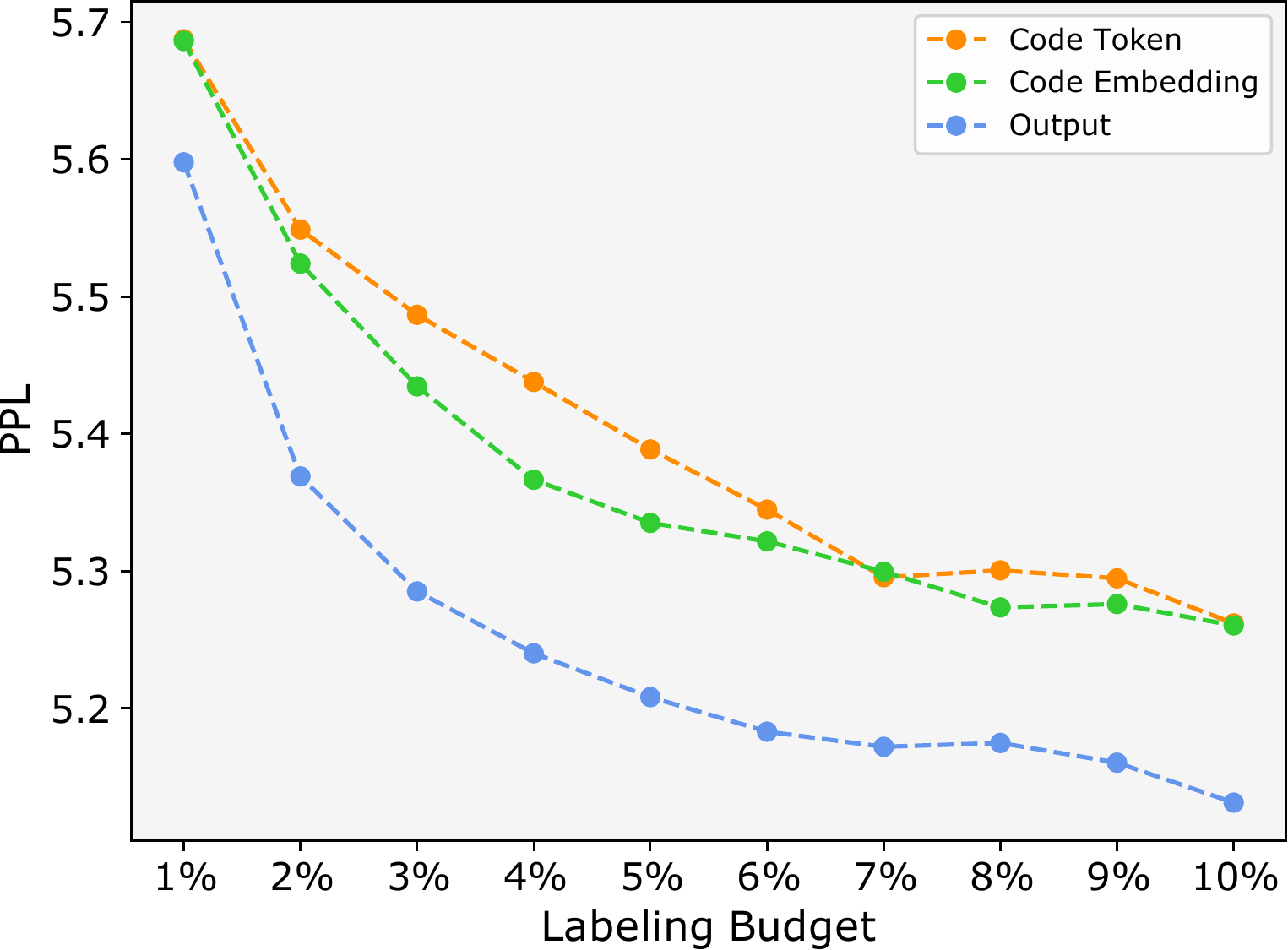}
    }
    \subfigure[Code Summarization-JS, BADGE]{
    \includegraphics[scale=0.27]{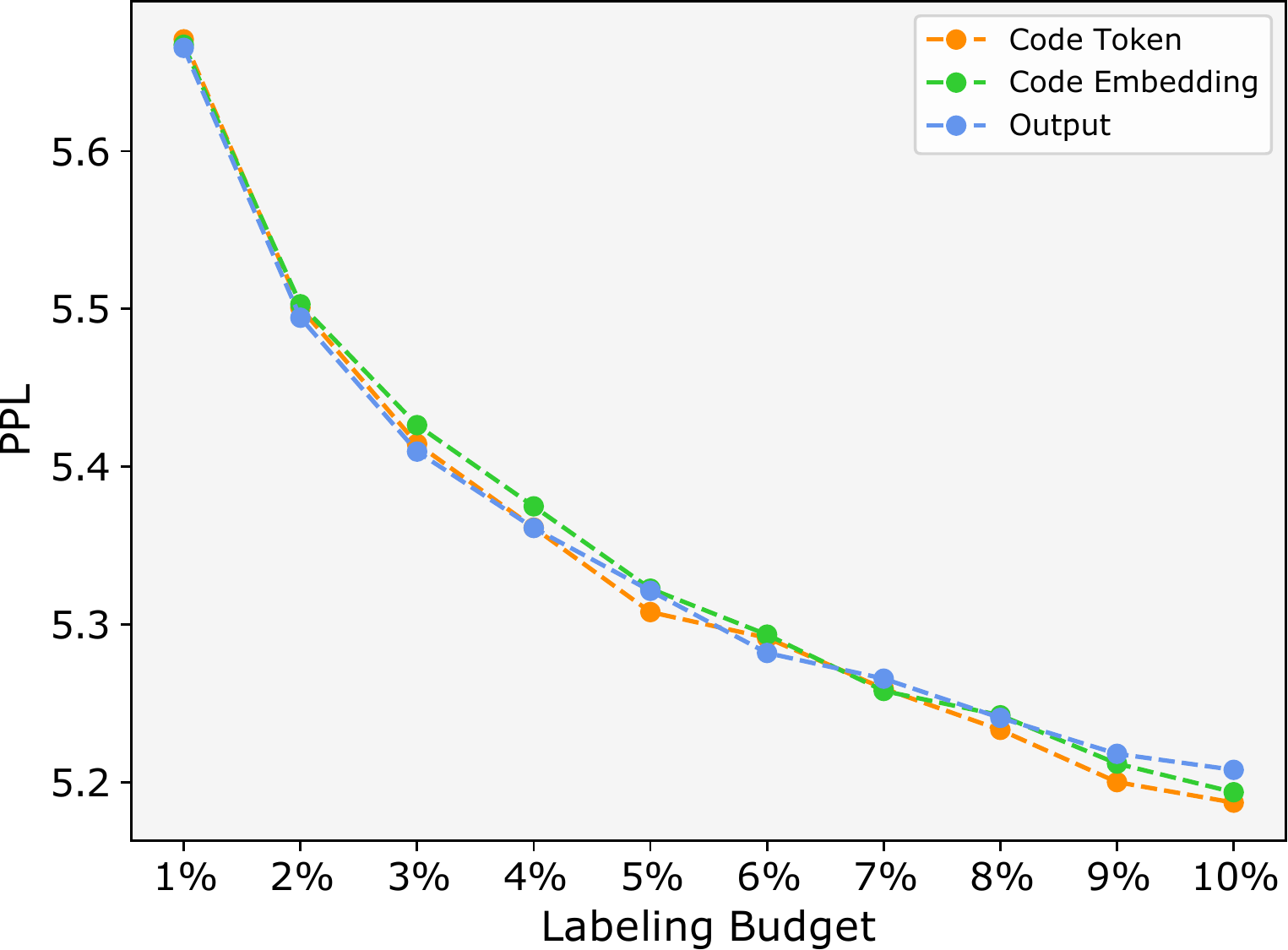}
    }
    \subfigure[Code Summarization-JS, Coreset]{
    \includegraphics[scale=0.27]{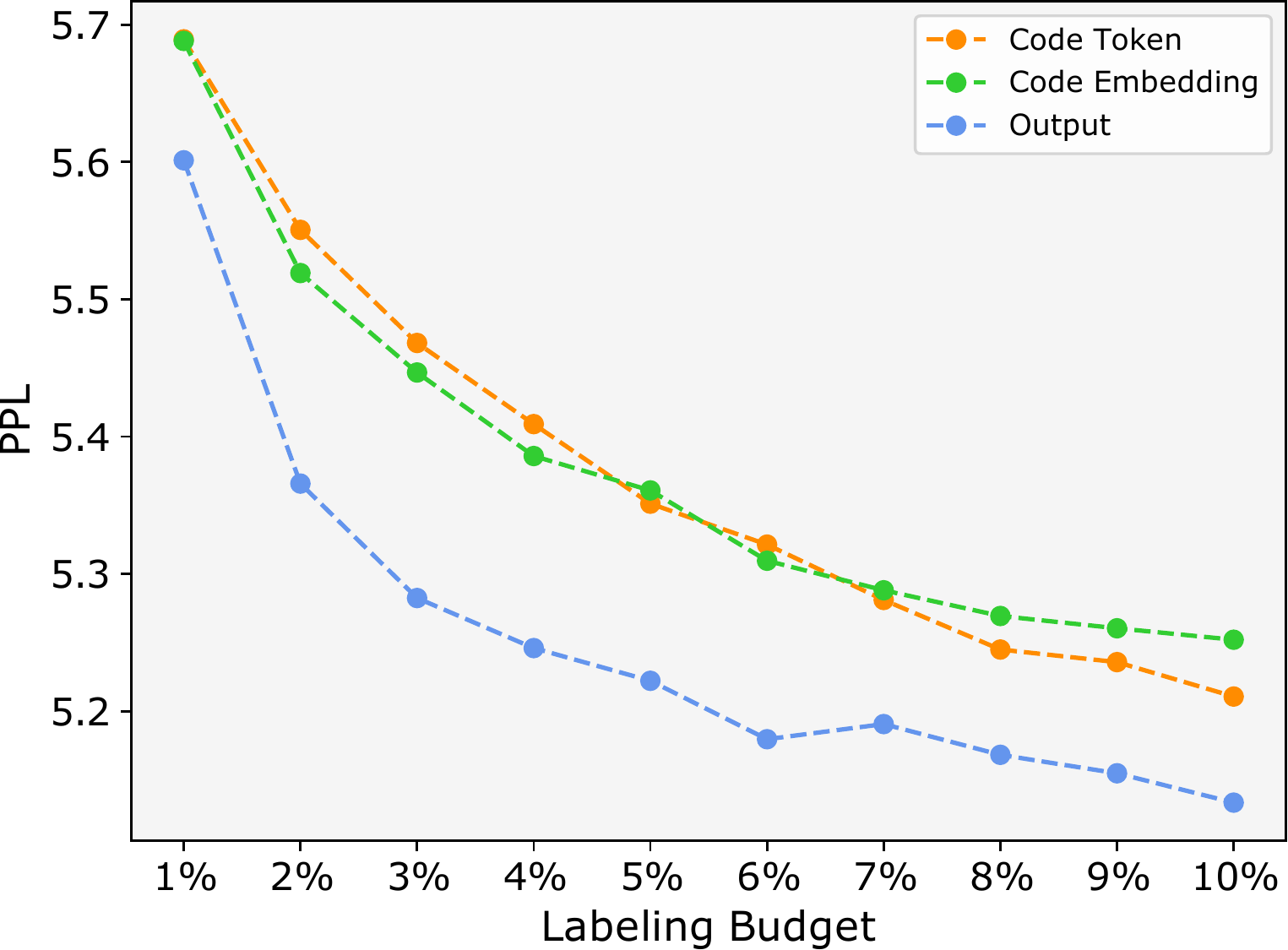}
    }
    \subfigure[Code Summarization-Ruby, KM]{
    \includegraphics[scale=0.27]{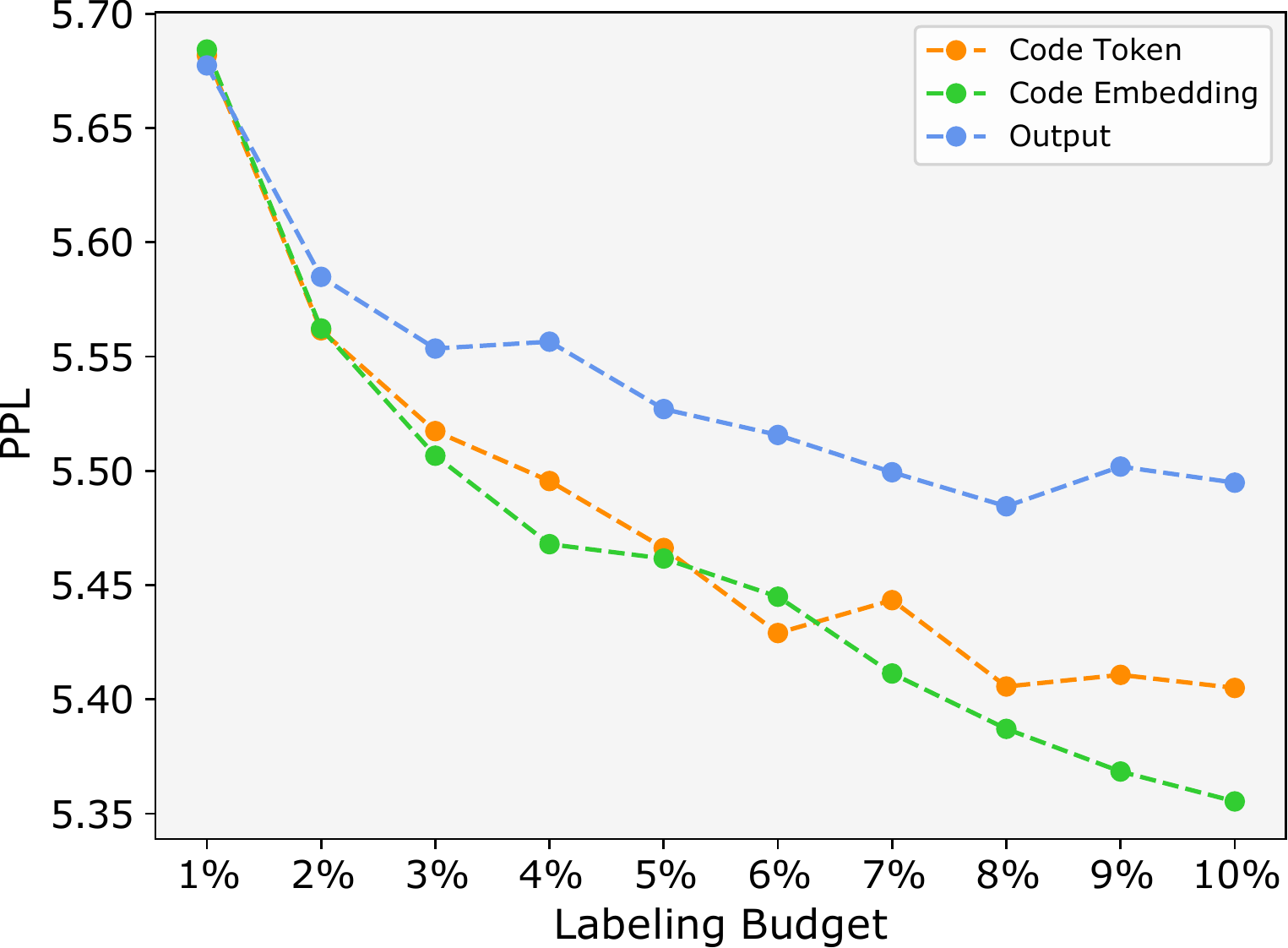}
    }
    \subfigure[Code Summarization-Ruby, KC]{
    \includegraphics[scale=0.27]{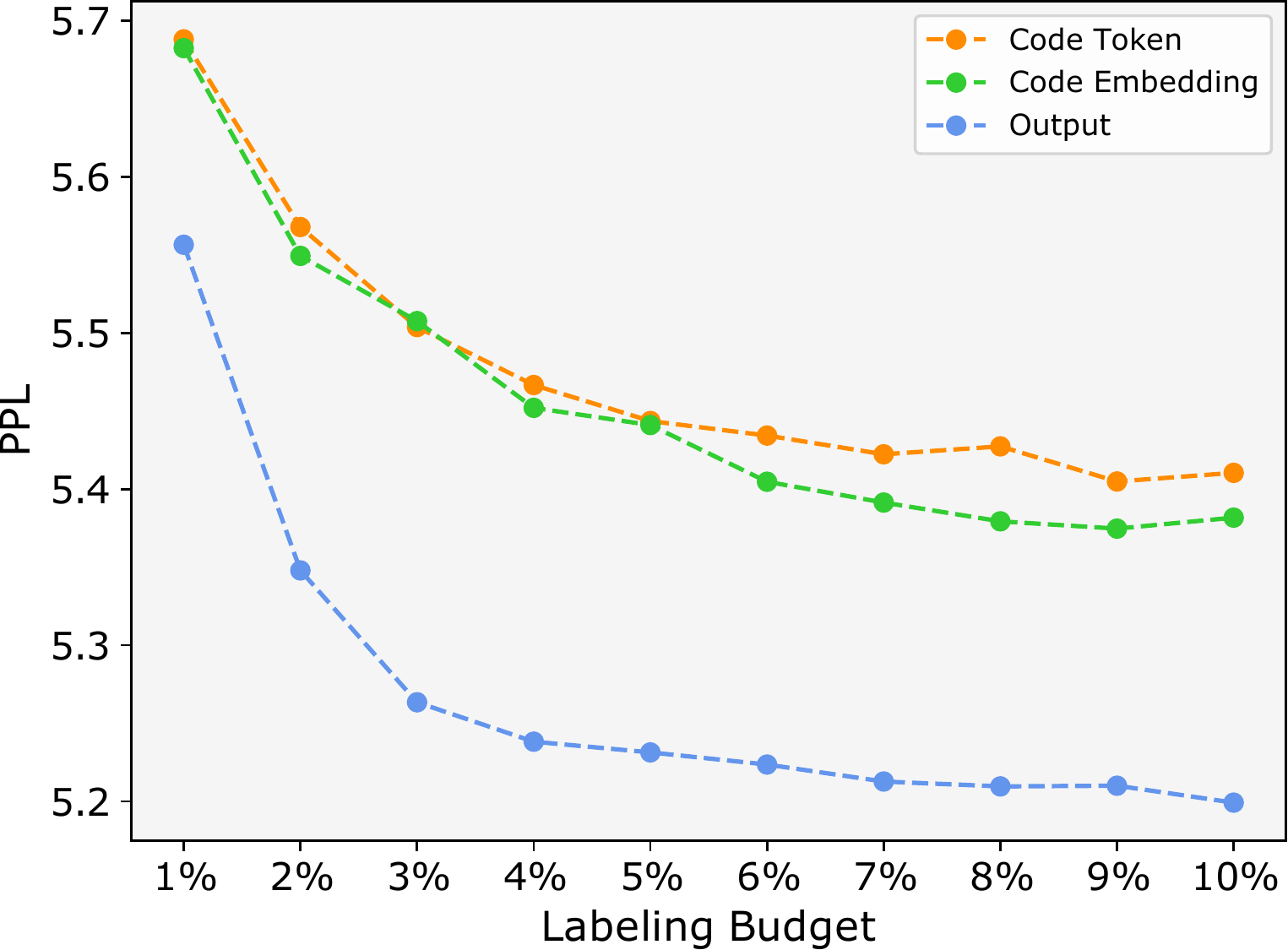}
    }
    \subfigure[Code Summarization-Ruby, BADGE]{
    \includegraphics[scale=0.27]{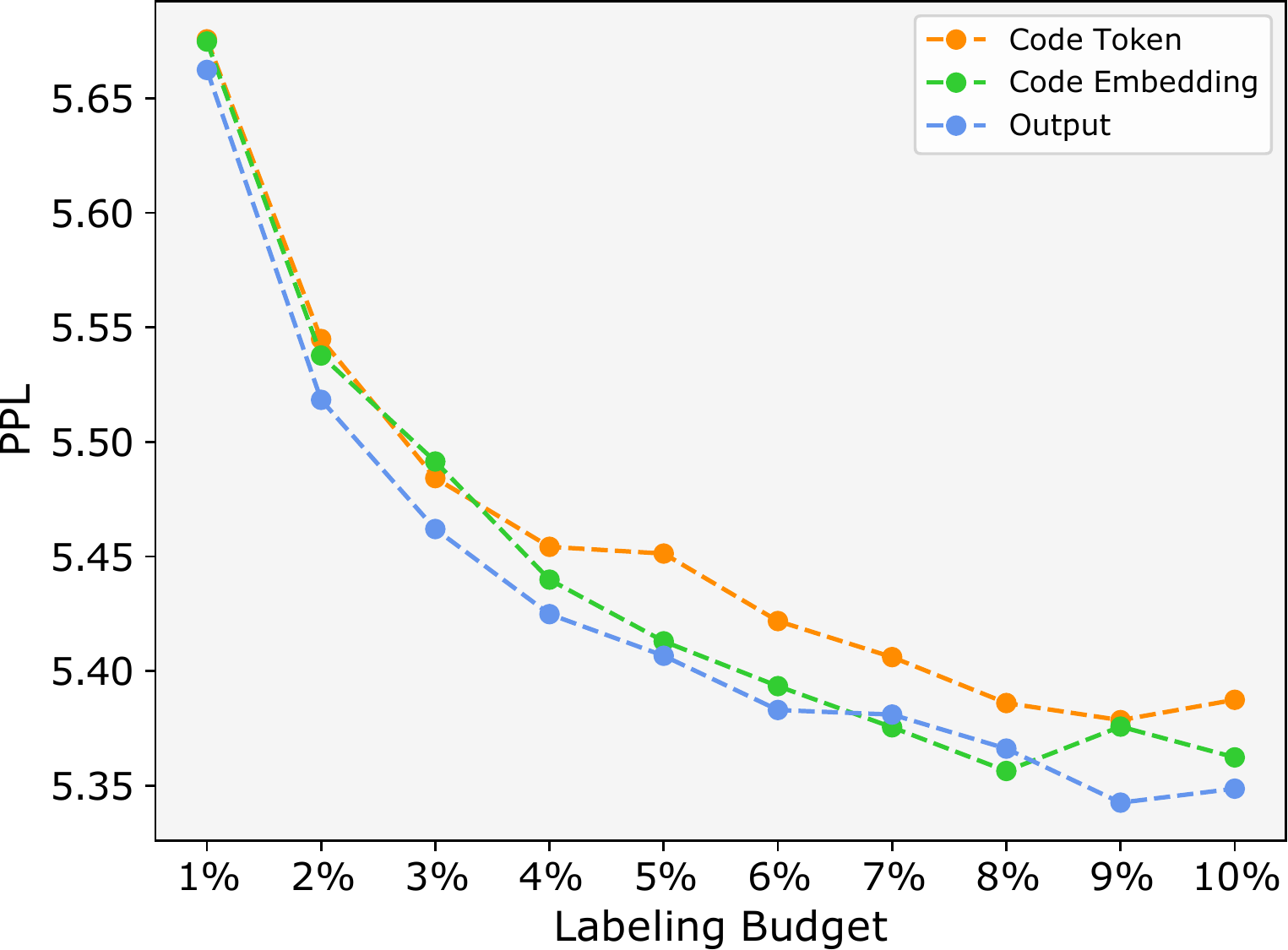}
    }
    \subfigure[Code Summarization-Ruby, Corest]{
    \includegraphics[scale=0.27]{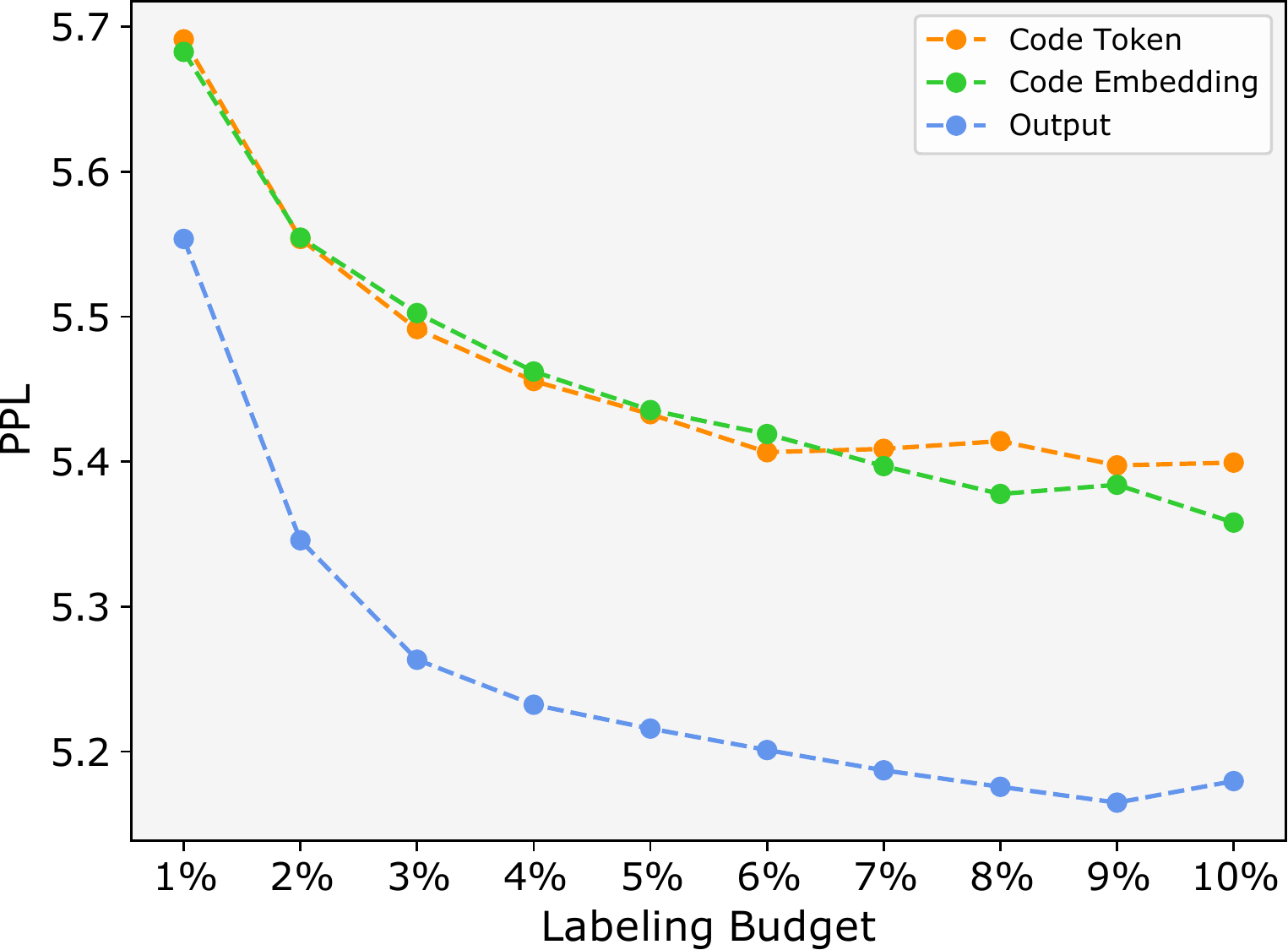}
    }
    \caption{Results of different feature-based active code learning. JS is short for JavaScript.}
    \label{fig:rq1}
\end{figure*}

First, we explore the features of clustering-based acquisition functions. Figure~\ref{fig:rq1} depicts the performance of models trained by different labeling budgets. Here, we only list the results of CodeBERT models and the whole results can be found on our project site. 

From the results, we can see that the same acquisition function could train models with significantly different performance depending on the features used. In particular, for the \textit{K-Means}, \textit{K-Center}, and \textit{Coreset} functions, the performance difference is substantial and should not be overlooked. In contrast, the \textit{BADGE} function is relatively stable compared to the others. Then, we analyze the most suitable features that should be used for each function for solving each code task. Table~\ref{table:rq1} displays the percentage of each function that produces models with the best performance across all labeling budgets.  Surprisingly, overall, the results demonstrate that the output vectors extracted from the models are superior features compared to code tokens and code embeddings. This suggests that active code learning should focus on output vectors rather than input vectors when conducting data selection, unlike active learning for image and text data.

\begin{table}[h]
\centering
\caption{Ratio of trained models achieving best results using different features.}
\resizebox{\columnwidth}{!}{
\begin{tabular}{lccccc}
\hline
 & \textbf{KM} & \textbf{KC} & \textbf{BADGE} & \textbf{Coreset} & \textbf{Average} \\ \hline
\multicolumn{6}{c}{Problem Classification} \\
Token & 0\% & 0\% & 0\% & 0\% & 0\% \\
Embedding & 10\% & 100\% & 40\% & 40\% & 47.5\% \\
Ouput & 90\% & 0\% & 60\% & 60\% & 52.5\% \\ \hline
\multicolumn{6}{c}{Clone Detection} \\
Token & 0\% & 40\% & 0\% & 0\% & 10\% \\
Embedding & 10\% & 60\% & 0\% & 40\% & 27.5\% \\
Ouput & 90\% & 0\% & 100\% & 60\% & 62.5\% \\ \hline
\multicolumn{6}{c}{Code Summarization} \\
Token & 50\% & 0\% & 20\% & 0\% & 17.5\% \\
Embedding & 45\% & 0\% & 15\% & 0\% & 15\% \\
Ouput & 5\% & 100\% & 65\% & 100\% & 67.5\% \\ \hline
\end{tabular}
}
\label{table:rq1}
\end{table}

Besides, we observe that the feature selection of some acquisition functions also depends on the types of downstream tasks. In both classification tasks, output vectors and code embeddings are the best choices for \textit{K-Means} and \textit{K-center}, respectively. However, In the non-classification code summarization task, the choices become code tokens and output vectors. On the other hand, output vectors are the most useful features for \textit{BADGE} and \textit{Coreset} regardless of the type of tasks. 

Based on the experimental results, we provide recommendations for the feature selection in clustering-based acquisition functions for active code learning.

\begin{itemize}[leftmargin=*]

\item \textbf{K-Means-C~(KM-C)}: use model output vectors~(code tokens) for classification~(non-classification) tasks.

\item \textbf{K-Center-C~(kC-C)}: use code embeddings~(model output vectors) for classification~(non-classification) tasks.

\item \textbf{BADGE-C}: use model output vectors for all code tasks. 

\item \textbf{Coreset-C}: use model output vectors for all code tasks. 

\end{itemize}

\noindent
\\
{\framebox{\parbox{0.96\linewidth}{
\textbf{Answer to RQ1}: Feature selection highly influences the effectiveness of clustering-based acquisition functions to perform active learning. Surprisingly, in most~(62.5\%) cases, active learning using output vectors extracted from the models can train a code model with better performance than using code tokens and code embeddings.}}}

\subsection{RQ2: Acquisition Function Comparison}

After studying the feature selection, we compare all acquisition functions on different code tasks. Here, we use our recommended features to build the clustering-based function. Table~\ref{table:rq2_classification} presents the results of classification tasks.

\begin{table*}[h]
\centering
\caption{Model performance of active learning trained models for two classification tasks. Values highlighted in red and blue indicate the best and second best.} 
\label{table:rq2_classification}
\begin{tabular}{lrrrrrr|rrrrrr}
\hline
 & \multicolumn{6}{c|}{\textbf{Problem Classification}} & \multicolumn{6}{c}{\textbf{Clone Detection}} \\
 & \multicolumn{3}{c}{CodeBERT} & \multicolumn{3}{c|}{GraphCodeBERT} & \multicolumn{3}{c}{CodeBERT} & \multicolumn{3}{c}{GraphCodeBERT} \\
 & 1\% & 5\% & 10\% & 1\% & 5\% & 10\% & 1\% & 5\% & 10\% & 1\% & 5\% & 10\% \\ \hline
\textbf{LC} & 29.61 & 52.90 & 59.14 & 33.99 & 56.93 & 61.06 & 80.08 & 83.51 & 90.85 & 79.90 & 81.73 & 90.12 \\
\textbf{Gini} & 43.96 & 90.95 & 97.75 & 44.31 & 93.02 & \cellcolor[HTML]{DAE8FC}98.04 & 85.45 & 96.58 & 96.74 & 84.45 & 96.41 & 97.18 \\
\textbf{Random} & 48.01 & 92.67 & 95.09 & 55.05 & 93.75 & 95.76 & 84.22 & 95.98 & 96.86 & 84.83 & 95.52 & 96.81 \\
\textbf{BALD} & 40.63 & \cellcolor[HTML]{DAE8FC}94.51 & \cellcolor[HTML]{DAE8FC}97.86 & 50.83 & 94.32 & 98.02 & 87.61 & 96.81 & 97.16 & 88.73 & 96.47 & 97.12 \\
\textbf{Entropy} & 42.67 & 90.05 & 97.67 & 44.50 & 92.21 & 97.99 & 86.16 & 96.32 & 96.89 & 85.28 & 96.89 & 97.17 \\
\textbf{Margin} & \cellcolor[HTML]{DAE8FC}51.22 & \cellcolor[HTML]{FFCCC9}95.41 & \cellcolor[HTML]{FFCCC9}97.91 & \cellcolor[HTML]{FFCCC9}57.76 & \cellcolor[HTML]{FFCCC9}96.33 & \cellcolor[HTML]{FFCCC9}98.16 & 86.38 & 96.82 & 96.92 & 84.92 & 97.02 & 97.17 \\
\textbf{CAL} & 29.99 & 54.51 & 64.99 & 34.04 & 57.68 & 65.33 & 79.57 & 91.10 & 95.14 & 77.60 & 88.45 & 92.06 \\
\textbf{KM-C} & \cellcolor[HTML]{FFCCC9}51.69 & 94.42 & 97.65 & \cellcolor[HTML]{DAE8FC}57.38 & 95.41 & 97.99 & \cellcolor[HTML]{DAE8FC}87.88 & \cellcolor[HTML]{DAE8FC}97.01 & 97.10 & 87.80 & 96.92 & \cellcolor[HTML]{DAE8FC}97.23 \\
\textbf{KC-C} & 47.84 & 94.01 & 97.73 & 56.18 & \cellcolor[HTML]{DAE8FC}95.57 & 97.33 & 87.48 & 95.07 & \cellcolor[HTML]{FFCCC9}97.79 & 85.77 & 96.85 & \cellcolor[HTML]{FFCCC9}97.62 \\
\textbf{BADGE-C} & 46.41 & 94.33 & 97.39 & 53.48 & 95.31 & 97.91 & \cellcolor[HTML]{FFCCC9}88.17 & \cellcolor[HTML]{FFCCC9}97.10 & \cellcolor[HTML]{DAE8FC}97.27 & \cellcolor[HTML]{DAE8FC}90.60 & \cellcolor[HTML]{FFCCC9}97.25 & 97.09 \\
\textbf{Coreset-C} & 47.96 & 94.23 & 97.66 & 56.35 & 95.48 & 97.24 & 86.45 & 96.95 & 96.96 & \cellcolor[HTML]{FFCCC9}92.10 & \cellcolor[HTML]{DAE8FC}97.19 & 97.06 \\ \hline
\end{tabular}
\end{table*}

For the multi-class classification task~(problem classification), we can see that output-uncertainty-based methods often achieve better results than the clustering-based methods. In which, \textit{Margin} which only uses the top-1 and top-2 probabilities of the outputs perform the best in 5 out of 6 cases. This phenomenon draws a similar conclusion to the previous work~\cite{weiss2022simple}, that simple methods perform well on active learning. However, considering the binary classification task~(clone detection), interestingly, the results show that the previous conclusion can not stand. In all cases, clustering-based methods outperform \textit{simple methods}~(output-uncertainty-based methods). Thus, the first conclusion we can draw is -- no acquisition functions that consistently perform better than others, and findings~\cite{weiss2022simple} from previous works can not be directly applied to active code learning.

\begin{table}[h]
\centering
\caption{PPL of active learning trained code summarization models with different training budgets. Values highlighted in red indicate the best.} 
\label{table:rq2_cs}
\resizebox{\columnwidth}{!}{
\begin{tabular}{lcccccc}
\hline
 & \multicolumn{6}{c}{\textbf{JavaScript}} \\
 & \multicolumn{3}{c}{CodeBERT} & \multicolumn{3}{c}{GraphCodeBERT} \\
 & 1\% & 5\% & 10\% & 1\% & 5\% & 10\% \\ \hline
Random & 5.6771 & 5.3407 & 5.2038 & 5.3806 & 5.1560 & 5.0938 \\
KM-C & 5.6840 & 5.3586 & 5.2219 & 5.3852 & 5.1589 & 5.0821 \\
KC-C & \cellcolor[HTML]{FFCCC9}5.5978 & \cellcolor[HTML]{FFCCC9}5.2082 & \cellcolor[HTML]{FFCCC9}5.1313 & \cellcolor[HTML]{FFCCC9}5.2883 & \cellcolor[HTML]{FFCCC9}5.0748 & \cellcolor[HTML]{FFCCC9}5.0178 \\
BADGE-C & 5.6654 & 5.3214 & 5.2079 & 5.3630 & 5.1379 & 5.0639 \\
Coreset-C & 5.6013 & 5.2222 & 5.1334 & 5.2912 & 5.0987 & 5.0346 \\ \hline
 & \multicolumn{6}{c}{\textbf{Ruby}} \\
 & \multicolumn{3}{c}{CodeBERT} & \multicolumn{3}{c}{GraphCodeBERT} \\
 & 1\% & 5\% & 10\% & 1\% & 5\% & 10\% \\ \hline
Random & 5.6671 & 5.4198 & 5.3662 & \cellcolor[HTML]{FFCCC9}5.3806 & \cellcolor[HTML]{FFCCC9}5.1560 & \cellcolor[HTML]{FFCCC9}5.0938 \\
KM-C & 5.6819 & 5.4662 & 5.4050 & 5.4038 & 5.1843 & 5.1100 \\
KC-C & 5.5565 & 5.2313 & 5.1992 & 5.4366 & 5.2179 & 5.1881 \\
BADGE-C & 5.6623 & 5.4067 & 5.3487 & 5.5942 & 5.3490 & 5.3015 \\
Coreset-C & \cellcolor[HTML]{FFCCC9}5.5536 & \cellcolor[HTML]{FFCCC9}5.2158 & \cellcolor[HTML]{FFCCC9}5.1796 & 5.4368 & 5.2120 & 5.1412 \\ \hline
\end{tabular}
}
\end{table}

\begin{table}[h]
\centering
\caption{BLEU score of active learning trained code summarization models with labeling budget 10\%. Values highlighted in red indicate the best.} 
\label{table:rq2_bleu}
\resizebox{\columnwidth}{!}{
\begin{tabular}{lcccc}
\hline
 & \multicolumn{2}{c}{\textbf{JavaScript}} & \multicolumn{2}{c}{\textbf{Ruby}} \\
 & CodeBERT & GraphCodeBERT & CodeBERT & GraphCodeBERT \\ \hline
Random & 9.62 & 10.10 & 10.36 & 11.60 \\
KM-C & 9.94 & 10.37 & 10.15 & 11.44 \\
KC-C & 9.96 & 10.32 & 10.46 & 11.55 \\
BADGE-C & 9.41 & 9.84 & \cellcolor[HTML]{FFCCC9}10.71 & 11.70 \\
Coreset-C & \cellcolor[HTML]{FFCCC9}10.09 & \cellcolor[HTML]{FFCCC9}10.42 & 10.45 & \cellcolor[HTML]{FFCCC9}11.71 \\ \hline
\end{tabular}
}
\end{table}

Then, move to the non-classification task~(two code summarization tasks), table~\ref{table:rq2_cs} and table~\ref{table:rq2_bleu} show the results. The first finding is that we will get different conclusions by using different evaluation metrics. For example, the PPL scores demonstrate that \textit{KC-C} is the best acquisition function while the BLEU scores suggest \textit{Coreset-C}. However, regardless of the evaluation metrics we used, the gap between the performance of active learning-trained models and the performance of models trained by using the entire data is big. For example, for JavaScript-CodeBERT, under 10\% labeling budget, the best PPL score and BLEU score we get are 5.1313 and 10.09, which are 33.28\% and 29.64\% lower than the 3.85 and 14.34 computed from the model trained by entire training data. These results are totally different from the ones drawn by the classification tasks that using 10\% data can train a model with similar and even better performance, e.g., for the clone detection task, models trained with 10\% ~(97.79\%) data have better performance than the models trained by entire training data ~(97.15\%). Therefore, we can say that active code learning in non-classification code tasks is still in a very early stage, the conclusions from classification tasks can not be migrated to non-classification tasks.

\noindent
\\
{\framebox{\parbox{0.96\linewidth}{
\textbf{Answer to RQ2}: In contrast to previous work on classification tasks~\cite{weiss2022simple}, our findings reveal that simple methods are ineffective for the binary code classification task -- clone detection. Clustering-based acquisition functions consistently outperform output-uncertainty-based functions in this task. In addition, active learning is ineffective for non-classification tasks such as code summarization, as the performance of models trained via active learning lags behind those trained using the entire dataset by at least 29.64\%.}}}

\section{Exploratory Study}
\label{sec:explore}

As discussed in Section~\ref{sec:bencj}, active code learning for non-classification code summarization tasks is still in an early stage. In this section, we tend to explore the potential directions to propose new effective acquisition functions and mainly focus on non-classification code tasks as existing acquisition functions are effective enough to produce good code models for classification tasks. Since clustering-based acquisition functions are the main techniques used for non-classification tasks, the main goal of our exploratory study is to explore ways to improve this type of acquisition function.

The key idea of clustering-based acquisition functions is to select a diverse set of data, each data sample in this set has big distances from the others. As a result, the calculation of the distance between each data is important in clustering-based functions. Thus, the straightforward way to improve the existing acquisition functions is to provide a precise way to measure the distance between code pairs~(or their features). Generally, in the existing acquisition functions, the distance is computed by the euclidean distance between two vectors that represent two programs. The main concern of this distance calculation is that vectors, e.g., code tokens, code embeddings, and output vectors, highly rely on code representation techniques or code models. It is difficult to say such computed distance can represent the real distance between two programs. 

Fortunately, recent research has proposed some evaluation metrics~\cite{eghbali2022crystalbleu, zhou2023codebertscore} for code generation. Roughly speaking, different from the existing distance methods, these metrics are specifically designed for code to measure the similarity between the machine-generated code snippets and the reference. The studies conducted by the original works show that there is a strong correction between the evaluation metrics and human preference. Inspired by these works, we propose to \textit{use the evaluation metrics as the distance methods} for the clustering-based acquisition functions. Ideally, this new distance should be more precise and the active code learning should be more effective.    

\subsection{Study Design}
To validate our conjectures, we empirically explore the following two problems:

\begin{itemize}[leftmargin=*]
\item \textit{Is there a correction between the distance of selected data to each other and the performance of trained models based on these data?} Through this study, we explore the probability of improving active code learning from the perspective of providing new distance methods for clustering-based acquisition functions.
\item \textit{Is there a connection between the existing distance methods and the code evaluation metrics?} Through this study, we explore whether our proposed distance methods are different from the old ones or not. The positive answer can demonstrate that evaluation-metric-based methods cannot be replaced by the existing distance methods. 
\end{itemize}

Addressing the first problem follows the following steps:

\textbf{Step 1} We prepare two groups of models, the first group contains the initialized models~(the same as the initialized models used in section~\ref{sec:bencj}) that represent models in the early stage of active learning, and the second group includes models that have already been trained using~5\% of training data that represent models at a late stage of active learning. In this way, we can see if the correlation we want to study holds in models with different performances. Note that we have prior knowledge of the data used to train the models.

\textbf{Step 2} We conduct active code learning by using \textit{Random} acquisition function for each group of models 100 times and record the 100 groups of selected data as well as the trained models for further analysis.

\textbf{Step 3} We measure the accuracy of the 100 trained models on the test data and calculate the average distance between the selected data samples in each group. Finally, we can get 100 accuracy values and corresponding 100 distance values. Here, the distance can be calculated by different methods.

\textbf{Step 4} We use Spearman's rank correlation coefficient to compute the correlation between the accuracy and the distance provided by \textbf{step 3}. 

For the second problem, we use different distance calculation methods in \textbf{Step 3} to obtain the distance scores and then use Spearman's rank correlation coefficient to compute the correlation between these distance 
methods.

\subsection{Setup}

We select one classification task~(problem Classification), and one non-classification task~(Code summarization for JavaScript programs) to conduct this exploratory study. For both tasks, we choose CodeBERT as our base model. For the distance calculation methods, we consider four in this study, cosine similarity as distance, Euclidean distance, BLEU score as distance, and CodeBERTScore~\cite{zhou2023codebertscore} as distance. Here, CodeBERTScore is a state-of-the-art evaluation metric for code generation. It uses pre-trained contextual embeddings to vectorize each token in the reference program and the generated program first. Then, it computes the pairwise cosine similarity between every embedded token in the reference and every encoded token in the generated code. Finally, the maximum similarity score in each row of the pairwise matrix is used to compute the final similarity of these two programs. Since the BLEU score and CodeBERTScore are computed based on the input level, we also compute the cosine distance and Euclidean distance based on the input embedding~(code embedding) for a more fair analysis.

\subsection{Results Analysis}

\begin{table}[]
\centering
\caption{Correlation between the selected data distance to each other and the performance of trained models based on these data. Each value represents the correlation coefficient. $Model_e$: model at the early stage of active learning. $Model_l$: at the late stage of active learning. Value with * indicates the P-value is less than 0.05.} 
\label{table:rq3_dis_acc}
\resizebox{\columnwidth}{!}{
\begin{tabular}{lcccc}
\hline
 & \multicolumn{2}{c}{\textbf{\begin{tabular}[c]{@{}c@{}}Classification\\ (Accuracy)\end{tabular}}} & \multicolumn{2}{c}{\textbf{\begin{tabular}[c]{@{}c@{}}Non-classification\\ (PPL)\end{tabular}}} \\
 & $Model_e$ & $Model_l$ & $Model_e$ & $Model_l$ \\ \hline
Cosine & 0.0589 & -0.2555* & 0.0705 & 0.1328 \\
Euclidean & -0.0773 & -0.2262* & 0.0868 & 0.0947 \\
BLEU & 0.0309 & -0.0589 & 0.0504 & -0.1699 \\
CodeBERTScore & 0.0463 & -0.0203 & 0.0128 & -0.1965* \\ \hline
\end{tabular}
}
\end{table}

Table~\ref{table:rq3_dis_acc} presents the results of the correlation between data distance and model accuracy. Surprisingly, the results indicate that regardless of the code tasks, when the model has a poor performance, i.e., at the early stage of active code learning, the trained model performance is not related to the diversity (the distance of data to each other) of used training data. However, for a model that was already trained on a few data and with a good performance, i.e., at the late stage of active code learning, the conclusion changed. Considering the classification task,  there is a weak correlation between the cosine and Euclidean distance with the accuracy of the models. That means clustering-based acquisition functions that use these two distance calculation methods are promising to train a model with high accuracy. As already shown in table~\ref{table:rq2_classification}, all these acquisition functions based on Euclidean distance achieve good performance in classification tasks. Considering the non-classification task, the results show this correlation becomes weaker, e.g., for cosine similarity, the correlation changes from -0.2555 to 0.1328 in \textit{$Model_l$}. This is also the reason that the existing acquisition functions do not work well on code summarization tasks and have no advantage over random selection as shown in table~\ref{table:rq2_cs} and table~\ref{table:rq2_bleu}. However, on the other hand, we can see there is a weak correlation between the evaluation scores-based distance and the accuracy of models which does not happen in our considered two classification tasks. The correlation result of CodeBERTScore on \textit{$Model_l$} is significant~(with a p-value less than 0.05). These results lead to a  promising direction of proposing new acquisition functions that use evaluation metrics as the distance calculation method for the code summarization task.

\noindent
\\
{\framebox{\parbox{0.96\linewidth}{
\textbf{Takeaway}: In non-classification code summarization tasks, our analysis shows that in the selected dataset, greater distances between data samples as calculated by evaluation-metrics-based distance methods lead to better model performance.
}}}\\

\begin{table}[]
\centering
\caption{Correlation between different distance calculation methods. Each value represents the correlation coefficient.  $Model_e$: model at the early stage of active learning. $Model_l$: model at the late stage of active learning. Value with * indicates the P-value is less than 0.05.} 
\label{table:rq3_dis_dis}
\resizebox{\columnwidth}{!}{
\begin{tabular}{lcccc}
\hline
 & \multicolumn{2}{c}{\textbf{Classification}} & \multicolumn{2}{c}{\textbf{Non-classification}} \\
 & $Model_e$ & $Model_l$ & $Model_e$ & $Model_l$ \\ \hline
Cosine-Euclidean & 0.9438* & 0.9155* & 0.9717* & 0.9769* \\
Cosine-BLEU & -0.0943 & -0.0356 & -0.1012 & -0.0111 \\
Cosine-CodeBERTScore & -0.2522 & -0.0838 & -0.4600* & -0.0030 \\
Euclidean-BLEU & -0.0963 & -0.0230 & -0.0856 & -0.0856 \\
Euclidean-CodeBERTScore & -0.2457 & -0.0615 & -0.4176* & -0.0038 \\
BLEU-CodeBERTScore & \multicolumn{1}{r}{0.6464*} & \multicolumn{1}{r}{0.6464*} & 0.2422 & 0.2422 \\ \hline
\end{tabular}
}
\end{table}

Table~\ref{table:rq3_dis_dis} presents the results of the correlation between different distance calculation methods. For models with low performance~(\textit{$Model_e$}), there is always a correlation between cosine similarity or euclidean distance with CodeBERSCore, which means CodeBERSCore is able to produce similar distance ranking of data to the existing distance methods in these models. Combining the conclusion from the last study, we can see that for \textit{$Model_e$}, all methods have a similar distance ranking of data, but this ranking is not connected to the performance of models. However, for \textit{$Model_l$}, we can see there is no correlation between cosine and euclidean to the BLEU and CodeBERTScore. That is the reason why cosine and euclidean~(BLEU and CodeBERTScore) correlate with the model performance while BLEU and CodeBERTScore~(cosine and euclidean) do not have this correlation in our considered classification~(non-classification) tasks.

\noindent
\\
{\framebox{\parbox{0.96\linewidth}{
\textbf{Takeaway}: Distance methods that are correlated with model performance produce significantly different rankings of data compared to methods that do not exhibit such a correlation.}}}

\subsection{Case Study}

According to our exploratory study, we know that evaluation metrics are promising to be used as distance methods in clustering-based acquisition functions. In this part, we conduct a case study to show if it can really help improve such acquisition functions. Specifically, we modify the distance calculation method in the \textit{Coreset} function from euclidean distance to BLEU and run experiments on code summarization tasks on Ruby. The reason we choose Ruby is that running evaluation metrics~(especially CodeBERTScore) is time-consuming, e.g., it takes more than one month to run an active learning experiment once on code summarization tasks of JavaScript since it contains 2 times more training data than Ruby. The reason for choosing \textit{Coreset} is that it performs the best in code summarization of Ruby as shown in table~\ref{table:rq2_cs} and table~\ref{table:rq2_bleu}. However, since CodeBERTScore does not support Ruby language now, we can only replace the original distance method in \textit{Coreset} with the BLEU metric. Note that we still use the euclidean distance between code embeddings to compute $Pairwise\ Distances$ which is the initial step of \textit{Coreset}. It is almost impossible to use the BLEU score here~(the time cost is monthly). Please refer to our project site for the details of the Coreset algorithm.






\begin{figure}[!h]
    \centering
    \subfigure[CodeBERT]{
    \includegraphics[scale=0.265]{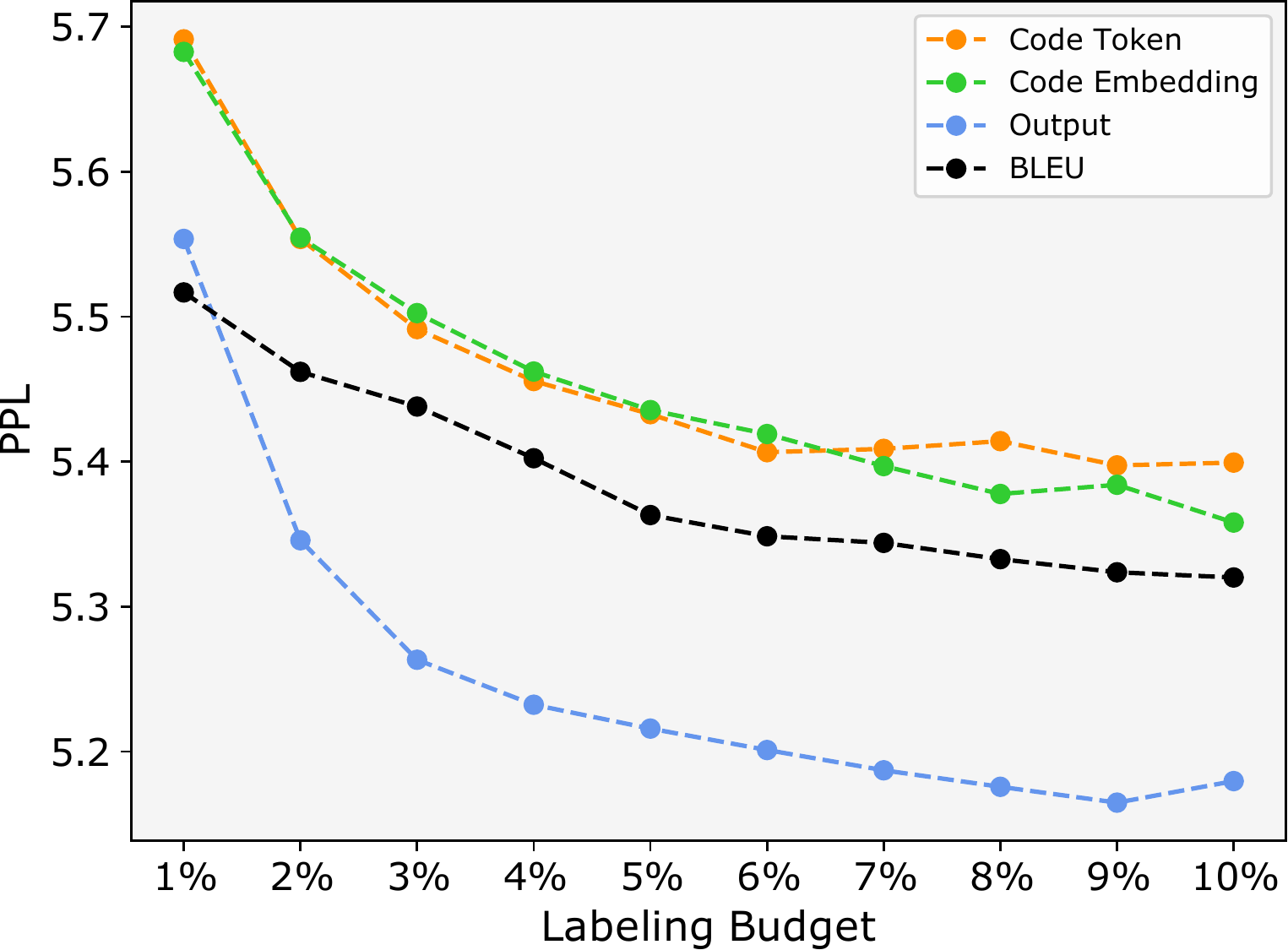}%
    }
    \subfigure[GraphCodeBERT]{
    \includegraphics[scale=0.265]{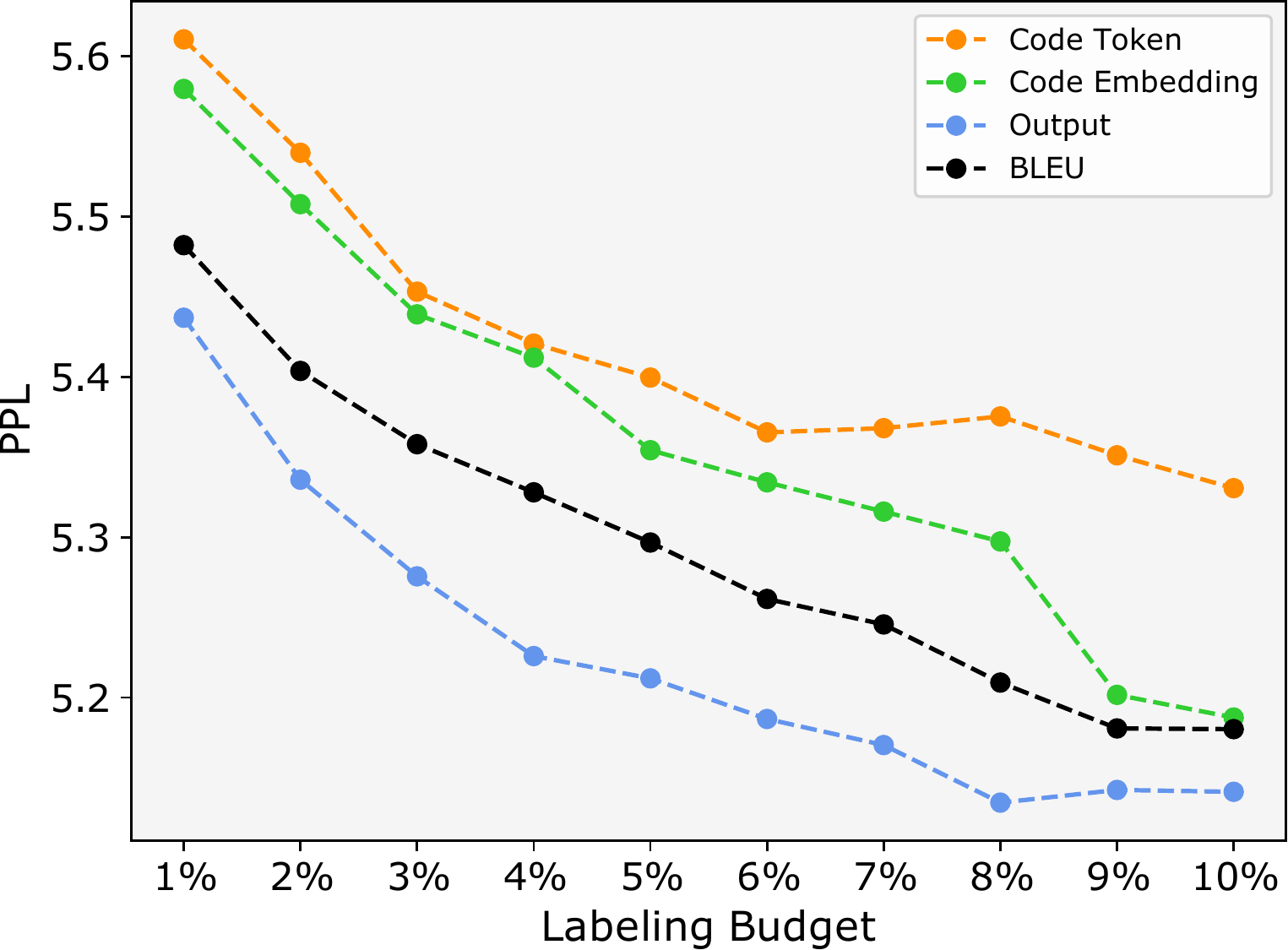}%

    }

    \caption{Active learning with Coreset acquisition functions. Code task: code summarization for Ruby. BLEU: replacing original Euclidean distance in Coreset to BLEU metric.}
    \label{fig:rq3}
\end{figure}

Figure~\ref{fig:rq3} depicts the results. We can see that \textit{Coreset} with a BLEU score performs significantly better than \textit{Coreset} with Euclidean distance calculated from code tokens and code embeddings. These results demonstrate the potential of using evaluation metrics as distance methods in active learning. However, using output vectors as clustering features which is also proposed by us is still the best choice which achieves the best results among all the cases. There is a big room to be improved in terms of the performance of trained models and how to propose a new acquisition function based on the evaluation metrics is still an open problem and our future research.


\section{Discussion}
\label{sec:discuss}
\subsection{The Importance of Active Code Learning}

Recently, large deep-learning models, especially foundation models like GPT-4~\cite{gpt4} have gained huge attention and achieved many state-of-the-art results in various application domains. This hot trend almost changes the research focus of ML4Code from designing new code model architectures or code representation techniques to how to reuse these foundation models for our specific code tasks. Generally, model reuse involves a fine-tuning step that further optimizes the model parameters and improves performance. As a result, active code learning becomes more and more important since it allows us to fine-tune the pre-trained models with a controllable human effort, i.e., budgets allocated for labeling the datasets used in fine-tuning. This technique provides opportunities for researchers and developers with limited resources to leverage and improve existing big models.

\subsection{Threat to Validity}

The \textbf{external threat} lies in our considered acquisition functions used for active learning, code tasks and datasets, and code models. For the acquisition functions, we collect 10 functions that are specifically proposed for active learning and already studied the most in recent works~\cite{hu2021towards, weiss2022simple}. Other functions such as  neural coverage methods are not considered since they are proposed for a different purpose. For code tasks and datasets, we consider both classification tasks that study important problems, problem classification, clone detection, and code summarization. For code models, we prepare two well-known code pre-trained models. Based on our open-source projects, other tasks, and models can be easily added to our benchmark. The \textbf{internal threat} can be the implementation of acquisition functions and code models. All implementations of acquisition functions are based on the existing active learning works~\cite{hu2021towards, li2022empirical, guo2022dre} and after carefully checking. The implementation of code models is also modified from the famous open source project~\cite{DBLP:journals/corr/abs-2102-04664}. The \textbf{construct threat} can be the configuration of active learning. Since this is the first work that studies active code learning, we follow our best practice to initialize the code models and set the labeling budgets. Besides, since we compare code models under the same labeling budgets, the comparison results are not affected by the configuration of active learning.

\section{Related Work}
\label{sec:related}
We review related works in two aspects: empirical study on active learning and empirical study on code learning.

\subsection{Empirical Study on Active Learning}

Since active learning plays an important role in efficient model training, multiple works~\cite{pereira2019empirical, settles2008analysis, yu2018finding, siddhant2018deep} conducted empirical studies on this topic. Ramirez-Loaiza~\emph{et al.}~\cite{ramirez2017active} compared different active learning methods using different measurements and concluded that improvements obtained by active learning for one performance measure often came at the expense of another measure. Heilbron~\emph{et al.}~\cite{heilbron2018annotate} studied active learning for a specific task, action localization. They found that using acquisition functions the select previously labeled data and combine them with the newly selected data is a more useful strategy of active learning for action localization. More recently, Hu~\emph{et al.}~\cite{hu2021towards} explored the limitations of active learning. They studied adversarial robustness and the ability to handle model compression of models trained by using active learning. The results showed that models trained with active learning can achieve competitive test accuracy but suffers from robustness and compression ability loss. Michael~\emph{et al.}~\cite{weiss2022simple} conducted a replicability study and showed that the simple active learning methods, e.g., \textit{DeepGini}, perform better in active learning than neuron coverage-based methods. The most recent work is \cite{li2022empirical} which did a very large empirical study of 19 active learning methods including both fully-supervised active learning methods and semi-supervised active learning methods. They concluded that semi-supervised learning benefits active learning and should be considered in proposing new methods.

Different from the existing works which mainly study image data or conventional text data, our work is the first one to focus on program data which is the key type of data in the software engineering field.

\subsection{Empirical Study on ML4Code}

ML4Code gained huge attention recently, researchers also conducted multiple empirical studies to explore the problems in the ML4Code field.  In the very early work~\cite{chirkova2021empirical}, Chirkova~\emph{et al.} empirically study the backbone model architecture of the later code models--Transformer on different code tasks, code completion, function naming, and bug
fixing. They found that Transformers can utilize syntactic information in source code to solve code tasks. Niu~\emph{et al.}~\cite{niu2023empirical} conducted a large-scale empirical study to compare pre-trained models of
source code. In total, we studied 19 pre-trained models and 13 software tasks and gave fine-grained suggestions for using and evaluating these models. Steenhoek~\emph{et al.}~\cite{steenhoek2022empirical} studied the deep learning models for a specific code task, vulnerability detection. Their experimental results showed that models trained by specific types of vulnerability perform better than models trained by all vulnerabilities. Increasing the size of the training dataset has limited benefits to the performance of trained models. Jiang~\emph{et al.}~\cite{jiang2023empirical} conducted the first study of pre-trained model reuse. Concretely, they interviewed practitioners from Hugging Face and identified the challenges of 
pre-trained model reuse, e.g., missing attributes, discrepancies, and model risks. Besides considering only the clean performance of models of code, Mastropaolo~\emph{et 
al.}~\cite{mastropaolo2023robustness} studied the robustness of GitHub Copilot which is a famous code generation model. They generated semantically equivalent natural language descriptions based on the seed description and checked if Copilot can generate the same code functions as the ones generated by the seed description. They found that almost half of the semantically equivalent but
different method descriptions result in different code recommendations which means the code generation models are not robust. Hu~\emph{et al.}~\cite{hu2023codes} provided shifted datasets and studied the generalization ability of code models under data distribution shift. They found that code models are not robust and can not handle distribution shifts properly. Finally, Nie~\emph{et al.}~\cite{nie-etal-2022-impact} studied the influence of used evaluation methods on code summarization and found that different evaluation methods lead to conflicting results which should gain attention for users during testing the code models. 

The above works focused on the clean performance, robustness, challenges of reuse, and evaluation methods of code models. However, non of them studied the important problem of how to reduce the labeling effort of training data and efficiently train the code models, which is our main purpose.

\section{Conclusion}
\label{sec:conclusion}

This paper introduced the first benchmark and an empirical study for the important yet unexplored problem -- active code learning. Our experimental results demonstrated that active code learning is effective to train code models with excepted high performance for classification tasks such as problem classification and clone detection. However, it is still in the early stage for non-classification tasks like code summarization. Besides, we conducted an exploratory study to show using evaluation metrics as distance calculation methods is a promising way for proposing new clustering-based acquisition methods. We believe that our benchmark as well as empirical studies will provide developers and researchers insights into efficiently reusing~(i.e., with little human effort) existing large pre-trained models for their specific code tasks.

\bibliographystyle{ACM-Reference-Format}
\bibliography{sample-base}

\end{document}